\documentclass[aps,prb,twocolumn,showpacs,10pt]{revtex4-1}
\usepackage{graphicx}
\usepackage{amssymb}
\usepackage{amsmath,color}
\usepackage{bbold}
\usepackage{comment}
\pdfoutput=1

\begin{document}

\def\pc{\frac{2\pi}{\Phi_0}}

\def\e{\varepsilon}
\def\f{\varphi}
\def\p{\partial}
\def\ba{\mathbf{a}}
\def\bA{\mathbf{A}}
\def\bb{\mathbf{b}}
\def\bB{\mathbf{B}}
\def\bD{\mathbf{D}}
\def\bd{\mathbf{d}}
\def\be{\mathbf{e}}
\def\bE{\mathbf{E}}
\def\bH{\mathbf{H}}
\def\bj{\mathbf{j}}
\def\bk{\mathbf{k}}
\def\bK{\mathbf{K}}
\def\bM{\mathbf{M}}
\def\bm{\mathbf{m}}
\def\bn{\mathbf{n}}
\def\bq{\mathbf{q}}
\def\bp{\mathbf{p}}
\def\bP{\mathbf{P}}
\def\br{\mathbf{r}}
\def\bR{\mathbf{R}}
\def\bS{\mathbf{S}}
\def\bu{\mathbf{u}}
\def\bv{\mathbf{v}}
\def\bV{\mathbf{V}}
\def\bw{\mathbf{w}}
\def\bx{\mathbf{x}}
\def\by{\mathbf{y}}
\def\bz{\mathbf{z}}
\def\bG{\mathbf{G}}
\def\bW{\mathbf{W}}
\def\Bn{\boldsymbol{\nabla}}
\def\Bo{\boldsymbol{\omega}}
\def\Br{\boldsymbol{\rho}}
\def\Bs{\boldsymbol{\hat{\sigma}}}
\def\bh{{\beta\hbar}}
\def\mA{\mathcal{A}}
\def\mB{\mathcal{B}}
\def\mD{\mathcal{D}}
\def\mF{\mathcal{F}}
\def\mG{\mathcal{G}}
\def\mH{\mathcal{H}}
\def\mI{\mathcal{I}}
\def\mL{\mathcal{L}}
\def\mO{\mathcal{O}}
\def\mP{\mathcal{P}}
\def\mT{\mathcal{T}}
\def\mZ{\mathcal{Z}}
\def\fr{\mathfrak{r}}
\def\ft{\mathfrak{t}}
\newcommand{\rf}[1]{(\ref{#1})}
\newcommand{\al}[1]{\begin{aligned}#1\end{aligned}}
\newcommand{\ar}[2]{\begin{array}{#1}#2\end{array}}
\newcommand{\eq}[1]{\begin{equation}#1\end{equation}}
\newcommand{\bra}[1]{\langle{#1}|}
\newcommand{\ket}[1]{|{#1}\rangle}
\newcommand{\av}[1]{\langle{#1}\rangle}
\newcommand{\AV}[1]{\left\langle{#1}\right\rangle}
\newcommand{\aav}[1]{\langle\langle{#1}\rangle\rangle}
\newcommand{\braket}[2]{\langle{#1}|{#2}\rangle}
\newcommand{\ff}[4]{\parbox{#1mm}{\begin{center}\begin{fmfgraph*}(#2,#3)#4\end{fmfgraph*}\end{center}}}

\title{Spin-dependent coupling between quantum dots and topological quantum wires}

\author{Silas Hoffman$^1$}
\author{Denis Chevallier$^1$}
\author{Daniel Loss$^1$}
\author{Jelena Klinovaja$^1$}
\affiliation{$^1$Department of Physics, University of Basel, Klingelbergstrasse 82, CH-4056 Basel, Switzerland}


\begin{abstract}
Considering Rashba quantum wires with a proximity-induced superconducting gap as physical realizations of Majorana fermions and quantum dots, we calculate the overlap of the Majorana wave functions with the local wave functions on the dot. We determine the spin-dependent tunneling amplitudes between these two localized states and show that we can tune into a fully spin polarized tunneling regime by changing the distance between dot and Majorana fermion. Upon directly applying this to the tunneling model Hamiltonian, we calculate the effective magnetic field on the quantum dot flanked by two Majorana fermions. The direction of the induced magnetic field on the dot depends on the occupation of the nonlocal fermion formed from the two Majorana end states which can be used as a readout for such a Majorana qubit.
\end{abstract}
 
\maketitle

\section{Introduction}
Majorana fermions\cite{mourikSCI12,dasNATP12,rokhinsonNATP12,dengNANOL12,finckPRL13,churchillPRB13,nadj-pergeSCI14, pawlakNPJ16} (MFs) are a promising candidate for topological quantum computation. Being spinless and chargeless particles, they are robust to disorder.\cite{kitaevAoP03,freedmanCiMP02, bravyiAoP02,freedmanBAMS03} However, these properties that make them a desirable element for information \textit{storage} make \textit{readout} problematic. Nonetheless, there have been several schemes for storage, manipulation, and readout of topological quantum computers using MFs modeled as a Kitaev chain,\cite{kitaevPU01} which is largely phenomenological.\cite{bravyiPRA06,hyartPRB13,leijnsePRL11,leijnsePRB12,pluggePRB16,
hoffmanPRB16,landauPRL16,karzigCM16,pluggeNJoP17}  A theoretical analysis of physically realized MFs for quantum information storage has yet to be rigorously studied; the details of which, as we show in this manuscript, are critical for quantum operations. 

Although there are several systems in which MFs have been proposed, perhaps the most readily accessible are quantum wires\cite{lutchynPRL10, oregPRL10} because: (1) there is potentially a large spin-orbit interaction (SOI), (2) advances in material science allow superconductivity to be easily induced by proximity, and (3) electrical gating allows the wire to be easily tuned in and out of the topological regime (see for instance Ref.~\onlinecite{albrechtNAT16}). When two ends of two quantum wires are brought close to each other, the two MFs at the ends form a nonlocal fermionic state which can be occupied or unoccupied. If a quantum dot,\cite{lossPRA98,kloeffelARCMP13} which can be electrically defined in experiments within the same quantum wire, is brought into proximity of these Majorana end states, the charge or spin coupling can be used to readout the parity of the quantum wire junction.\cite{hoffmanPRB16}

\begin{figure}
\includegraphics[width=1\linewidth]{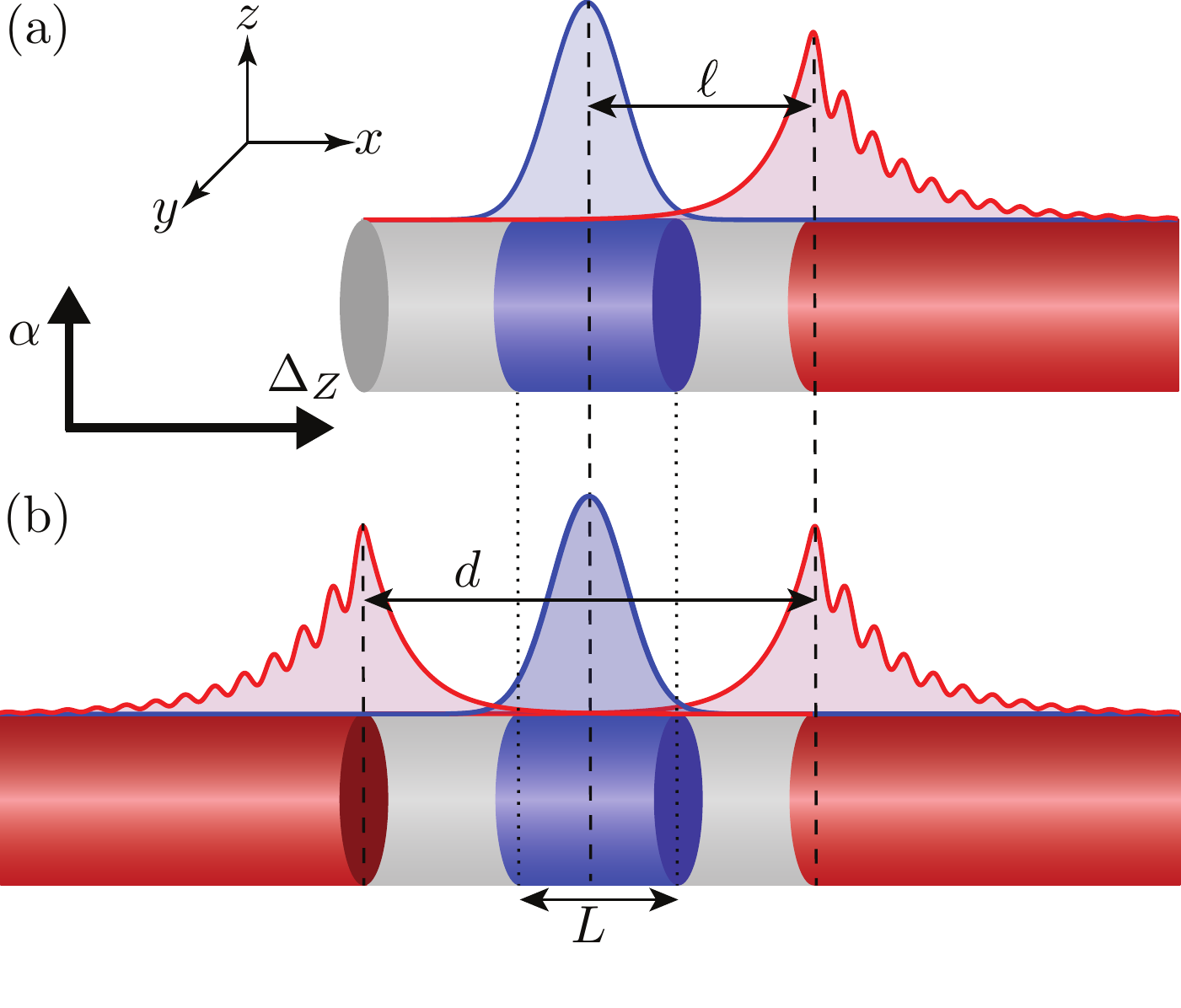}
\caption{A quantum wire with applied magnetic field along the longitudinal ($x$) axis and spin-orbit vector along the $z$ axis in which the boundary between a topological section (red) and nontopological section (grey) supports a MF. In the upper panel, a quantum dot (blue) of size $L$ is defined, within the nontopological section, at a distance $\ell$ from the topological section. The second setup (lower panel) is identical to the first with an additional topological section which ends a distance $d-\ell$ from the quantum dot center. The red and blue curves are schematically the probability amplitudes of the MF  and quantum dot wave functions, respectively.}
\label{setup1}
\end{figure}

In this manuscript, we study MFs formed in a quantum wire with proximity-induced superconductivity near a quantum dot,\cite{vernekPRB14,dengSCI16,riccoPRB16,dessottiPRB16,schradeCM16,xuCM16,guessiCM17,pradaCM17} which is also formed inside a quantum wire, all of which are subject to an applied magnetic field perpendicular to the SOI. This allows us to calculate the spin-dependent tunneling amplitudes between dot and MFs, which depend on the size of the dot and the distance from the dot to the MFs. By changing the relative position between dot and  MFs, one can tune between spin-independent and fully polarized tunneling for typical parameters. In the presence of two MFs, the spin-dependent tunneling induces an effective magnetic field on the dot which changes direction when the occupation of the nonlocal fermion, formed from the two MFs, changes parity. Thus, this setup allows the read out of MF qubits via reading out the spin of the electron on the quantum dot.\cite{hoffmanPRB16} 

We organize the manuscript as follows: in Sec. \ref{model}, we describe the quantum wire that hosts two MFs and a quantum dot. In Sec.~\ref{overlap}, we calculate the overlap of the MF and quantum dot wave functions, and thus the spin-dependent tunneling, for which we obtain simple analytic expressions
in a suitable limit. Using these results, in Sec.~\ref{readout}, we consider two MFs coupled to the dot and calculate the effective magnetic field when the complex fermion state formed from the MFs is occupied or unoccupied. In Sec.~\ref{nums}, we numerically calculate the spin-dependent coupling and the effective magnetic field on the dot
using a tight binding model. We conclude in the final section with a summary of our results and their implications on proposed quantum dot-MF computational schemes.

\section{Model}\label{model}

We consider a quantum wire in proximity to a conventional superconductor, so superconducting pairing is induced, and a magnetic field along the longitudinal axis which is perpendicular to the spin-orbit direction (see Fig.~\ref{setup1}). There is full spatial control of the chemical potential over the wire so that the right section and left section are tuned into the topological and nontopological regimes, respectively, with appropriate gating; at the intersection resides a MF. Within the nontopological section, appropriate gates define a quantum dot which supports a localized wave function. For sufficiently large barrier between the MF and quantum dot, which we assume in the following, the MF wave function can be solved independently from the quantum dot wave functions within the same quantum wire.

\subsection{Majorana Fermion}
\label{secQW}
To find the MF wave function, we consider the two sections in the quantum wire, topological ($\nu=t$) and nontopological ($\nu=n$), to be kept at different chemical potentials, $\mu_{n}$ and $\mu_t$,  and whose interface is at $x=\ell$. The Hamiltonian of this system is
\begin{equation}
H=\int dx\ \Psi^\dagger(x)\mathcal H\Psi(x)\,,
\end{equation}
where $\mathcal H=\mathcal H_0+\mathcal H_{SO}+\mathcal H_s+\mathcal H_{Z}$ is composed of a kinetic, $\mathcal H_0=-\eta_3[\hbar^2 \partial_x^2/2m+\mu(x)]$, SOI interaction, $\mathcal H_{SO}=-i\alpha\sigma_3\partial_x$, Zeeman, $\mathcal H_Z=\Delta_Z\sigma_1\eta_3$, and superconducting pairing, $\mathcal H_s=\Delta_{s}\sigma_2\eta_2$, terms, correspondingly. 
Here, $\Psi(x)=[\psi_\uparrow(x),~\psi_\downarrow(x)~\psi^\dagger_\uparrow(x),~\psi^\dagger_\downarrow(x)]^T$ is the Nambu spinor in the quantum wire, $\mu(x)=\mu_t\Theta(x-\ell)+\mu_{n}\Theta(\ell-x)$, $\alpha$ is the SOI constant, $\Delta_Z$ is the Zeeman splitting due to the applied magnetic field, $\Delta_s$ is the proximity induced superconducting gap, and $\psi_\sigma(x)$ [$\psi^\dagger_\sigma(x)$] annihilates (creates) an electron of spin $\sigma=\uparrow,\downarrow$ quantized along the $z$ axis at position $x$. The Pauli matrices $\sigma_i$ and $\eta_i$ act in spin and particle-hole space, respectively. The condition $\Delta_Z^2>\mu_{t}^2+\Delta_s^2$ is necessary to be in the topological phase.\cite{satoPRB09,lutchynPRL10,oregPRL10,klinovajaPRB12} 

In the following, we consider the SOI energy to be large compared to the magnetic field ($E_{SO}=m\alpha^2/\hbar^2\gg \Delta_Z$) and the superconducting gap ($E_{SO}\gg\Delta_s$). We tune the right section of the wire into the topological regime by fixing the chemical potential to zero, $\mu_{t}=0$, and applying a large enough magnetic field such that the Zeeman splitting is larger than the superconducting gap, {\it i.e.}, $\Delta_Z>\Delta_s$. We consider two ways in which the left section can be driven into the nontopological regime: (1) chemical potential is small compared to the SOI energy but large enough so that $\Delta_Z^2<\Delta_s^2+\mu_{n}^2$, or (2) a fully depleted wire, $-\mu_{n}\gg E_{SO}\gg\Delta_Z,\,\Delta_s$, which is insulating in the normal phase. Although the second regime presents a more physical experimental realization\cite{lossPRA98}, 
 we are unable to analytically progress beyond the zero bulk solutions to the Hamiltonian, i.e. we cannot satisfy differentiability of the MF wave functions at the boundary (Appendix \ref{wave_analytic}). Therefore, we consider the former case which yields simple analytic results that are instructive in guiding the numerical methods used to solve the system in the latter regime (see below).

When the chemical potential is much smaller than the SOI energy, it is standard to go to the rotating frame of reference,\cite{brauneckerPRB10} dropping fast oscillating terms, to obtain a linearized Hamiltonian \cite{klinovajaPRB12}.  Rotating back to the lab frame, the zero energy eigenfunctions are given by 
\begin{widetext}
\begin{align}
\Phi_1^\nu&=\left( \begin{array}{c} -i~\textrm{sgn}(\Delta_s-\sqrt{\Delta_Z^2-\mu^2_\nu})e^{i\varphi_\nu/2}\\e^{-i\varphi_\nu/2}\\i~\textrm{sgn}(\Delta_s-\sqrt{\Delta_Z^2-\mu^2_\nu})e^{-i\varphi_\nu/2}\\e^{i\varphi_\nu/2}\end{array} \right) e^{-\kappa^\nu_1 (x-\ell)}, \,\, \,\,\, \,
\Phi_2^\nu=\left( \begin{array}{c} e^{-i\varphi_\nu/2}\\-i e^{i\varphi_\nu/2} \\e^{i\varphi_\nu/2}\\i e^{-i\varphi_\nu/2} \end{array} \right) e^{-\kappa^\nu_2 (x-\ell)}\,,\nonumber\\
\Phi_3^\nu&=\left( \begin{array}{c} ie^{2i k_{SO} (x-\ell)}\\e^{-2i k_{SO} (x-\ell)}\\-ie^{-2i k_{SO} (x-\ell)}\\e^{2i k_{SO} (x-\ell)}\end{array} \right)
 e^{-\kappa^\nu (x-\ell)}\,, \,\, \,\,\, \, \Phi_4^\nu=\left( \begin{array}{c} e^{2i k_{SO} (x-\ell)}\\i e^{-2i k_{SO} (x-\ell)}\\e^{-2i k_{SO} (x-\ell)}\\-i e^{2i k_{SO} (x-\ell)}\end{array} \right)
 e^{-\kappa^\nu (x-\ell)}\,,
 \label{WFs}
\end{align}
\end{widetext}
where $\kappa^\nu_1=\pm|\Delta_s-\sqrt{\Delta_Z^2-\mu^2_\nu}|/\alpha$, $\kappa^\nu_2=\pm(\Delta_s+\sqrt{\Delta_Z^2-\mu^2_\nu})/\alpha$, $\kappa^\nu=\pm\Delta_s/\alpha$ and $\sin\varphi_\nu=\mu_\nu/\Delta_Z$ for which we require $\Delta_Z\geq\mu_\nu$. The $\pm$ in the real part of the exponentials refers to $\nu=t,n$, respectively, and $k_{SO}=m\alpha/\hbar^2$ is the SOI wave vector. Here, we have neglected terms $\mu_n/\alpha\ll k_{SO}$ in the wave functions $\Phi^n_3$ and $\Phi^n_4$ which renormalize the oscillations due to a shift in the Fermi points.

The wave functions in Eq.~(\ref{WFs}) are zero-energy eigenvectors of the Hamiltonian but do not individually satisfy the boundary conditions.  The MF wave function satisfying continuity and differentiability at the boundary is $\Phi_M=\Theta(x-\ell)\Phi^{t}+\Theta(\ell-x)\Phi^{n}$, where
\begin{align}
\Phi^{t}&=\mathcal{N}\left(\Phi_1^{t}-\frac{\kappa^{t}_1+\kappa^{t}}{2\kappa^{t}}\Phi_3^{t}+\frac{k_{SO}}{\kappa^{t}}\Phi_4^{t}\right)\,,\nonumber\\
\Phi^{n}& =\mathcal{N}\left(\frac{\kappa^{t}-\kappa^{t}_1}{2\kappa^{t}}\Phi_3^{n}+\frac{k_{SO}}{\kappa^{t}}\Phi_4^{n}\right)\,,
\label{maj_wav}
\end{align}
and where $\mathcal{N}$ is an overall normalization factor.
The probability amplitude of the MF, $|\Phi_M|^2$, on the topological section oscillates with half the spin-orbit wavelength, $\lambda_{SO}/2=\pi/k_{SO}$ and has two decay lengths given by the superconducting gap $\Delta_s$, and the induced gap, $|\Delta_Z-\Delta_s|$. On the nontopological section, although the components of the MF wave function oscillate with the same $2k_{SO}$-periodicity, the probability amplitude is a monotonically decreasing exponential with decay length $\Delta_s/\alpha$ [see Fig.~\ref{setup1} (red part)]. The shape of MF wave functions could be mapped experimentally using the STM techniques \cite{nadj-pergeSCI14, rubyPRL15,pawlakNPJ16,chevallierPRB16}.

\subsection{Quantum Dot}

There are two characteristic regimes in which one can create the quantum dot, (1) when the dot size is smaller than the spin-orbit wavelength, $\lambda_{SO}$, and (2) when it is larger. In the first case, the SOI term can be neglected while in the second case the spin components of the wave function oscillate on the wavevector $k_{SO}$.\cite{trifPRB08} 
Experimentally, the spatial profile of the superconductivity and gates between the quantum dot and MF wave functions could be complicated. However, we expect this to contribute only a spin \textit{independent} factor to the tunneling, which can be absorbed as a phenomenological parameter. In order to simplify the calculation, we consider a fully depleted section of the wire so that we can ignore the superconducting correlations on the dot. This allows us to analytically calculate the dot wave function and thus the spin \textit{dependent} tunneling which is the focus of the manuscript. 

\textit{Small Dot.}-- In the first case, the dot is described by 
\begin{equation}
H^s_D=\int dx \Psi^\dagger(x)[\mathcal H_0+\mathcal H_Z+\mathcal V(x)] \Psi(x)\,,
\label{dotS}
\end{equation} 
where $\mathcal V(x)$ is a confining potential defining the dot. For a parabolic confinement, $\mathcal V(x)=m \omega_0^2x^2/2-\mu_d$, where $\mu_d$ is a dot plunger potential, the lowest energy eigenvectors of $H^s_D$ are $X^i(x)=(1/4\pi L^2)^{1/4}\exp(-x^2/2L^2)\chi^i$, where $L=\sqrt{\hbar/m\omega_0}$ and $(\chi^1)^T=(1,1,0,0)$, $(\chi^2)^T=(-1,1,0,0)$, $(\chi^3)^T=(0,0,1,1)$, $(\chi^4)^T=(0,0,-1,1)$, 
with eigenenergies $\epsilon_0-\Delta_Z$, $\epsilon_0+\Delta_Z$, $-\epsilon_0-\Delta_Z$, and $-\epsilon_0+\Delta_Z$, respectively, and $\epsilon_0=\hbar\omega_0/2-\mu_d$.

\textit{Large Dot.}-- In the second case, the Hamiltonian is 
\begin{equation}
 H^l_D=\int dx \Psi^\dagger(x)[\mathcal H_0+\mathcal H_{SO}+\mathcal H_Z +\mathcal V(x)] \Psi(x)\,,
\end{equation}
where $\mathcal H_{SO}$ is the SOI as given in Sec.~\ref{secQW} and $\mathcal V(x)$ is a parabolic confinement as in the case of the small dot. Treating the magnetic field perturbatively as compared to the other energies on the dot, one may show that the Hamiltonian reduces to Eq.~(\ref{dotS}) in the rotating frame of reference with an exponentially renormalized magnetic field according to ratio of the dot size and SOI length, $\bar\Delta_Z=\Delta_Z e^{- k_{SO}L}$.\cite{trifPRB08}
The eigenvectors are
$Y^i(x)=(1/4\pi L^2)^{1/4}\exp(-x^2/2L^2)\zeta^i$ where $(\zeta^1)^T=(e^{ i k_{SO}x},e^{- i k_{SO}x},0,0)$, $(\zeta^2)^T=(-e^{ i k_{SO}x},~e^{- i k_{SO}x},~0,~0)$, $(\zeta^3)^T=(0,0,e^{-i k_{SO}x},e^{i k_{SO}x})$, $(\zeta^4)^T=(0,0,-e^{-i k_{SO}x},e^{i k_{SO}x})$,
with eigenenergies $\epsilon_0-\bar\Delta_Z$, $\epsilon_0+\bar\Delta_Z$, $-\epsilon_0-\bar\Delta_Z$, and $-\epsilon_0+\bar\Delta_Z$, respectively.

\begin{figure}[b]
\includegraphics[width=1\linewidth]{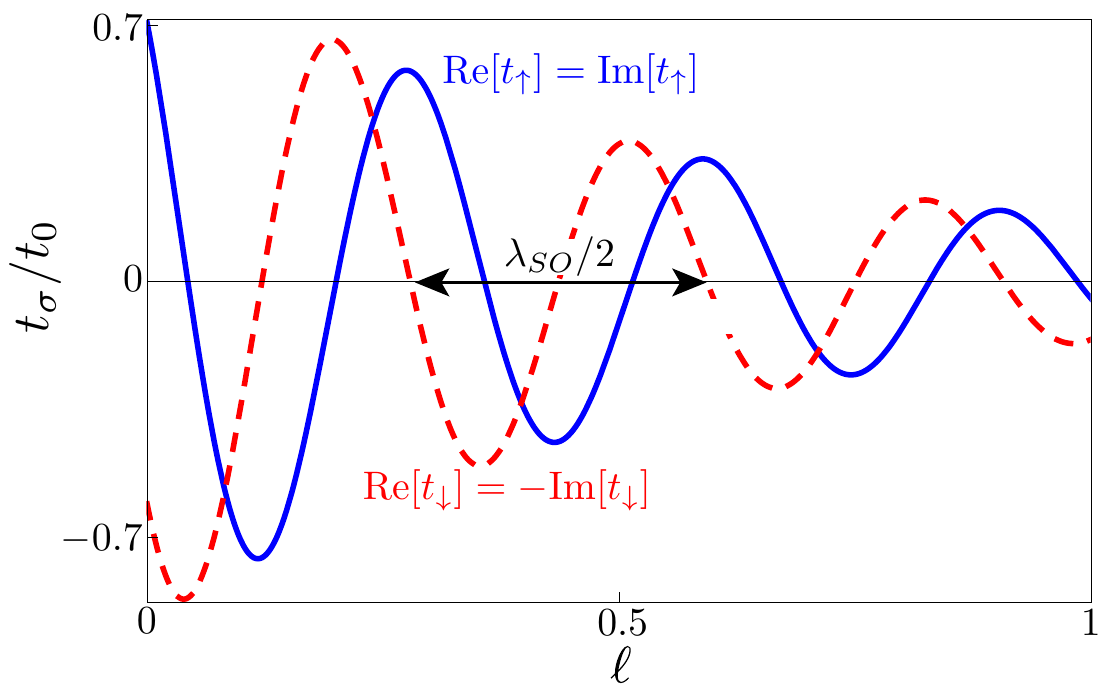}
\caption{The real and imaginary components of spin-dependent tunneling amplitudes $t_\uparrow$ (blue solid) and $t_\downarrow$ (red dashed) for  a small quantum dot, $L=0.01$, $k_{SO}=10$,  and $\kappa^{t}=\sqrt{3}$. Both $t_\uparrow$ and $t_\downarrow$ oscillate with periodicity of $\lambda_{SO}/2$ but with a relative phase difference of $\pi/2$.}
\label{t_small}
\end{figure}

\section{spin-dependent Tunneling}
\label{overlap}
It is now straightforward to evaluate the tunneling amplitudes between the MF and quantum dot states which are proportional to the overlap of the two corresponding wave functions,
\begin{align}
t_\uparrow&=\bar t_0\int dx (X^1)^\dagger\cdot  \Phi_M=\left(\int dx (X^3)^\dagger\cdot  \Phi_M\right)^*\,,\nonumber\\
t_\downarrow&=\bar t_0\int dx (X^2)^\dagger\cdot  \Phi_M=\left(\int dx (X^4)^\dagger\cdot  \Phi_M\right)^*\,,\nonumber\\
\end{align} 
where $\bar t_0$ is a phenomenological constant that is defined according to the potential profile separating the dot and MF.  The tunneling Hamiltonian between the dot and MF is\cite{leijnsePRL11,hoffmanPRB16}
\begin{equation}
H_T=\sum_\sigma t_\sigma d_\sigma^\dagger \gamma +\textrm{H.c.}\,,
\end{equation}
where $\gamma$ is the MF operator and $d_\sigma^\dagger$ creates an electron with spin $\sigma$, quantized along the axis of the magnetic field ($x$ axis).

In the limit that $k_{SO} L \ll 1$, which implies that $L$ is much smaller than the MF decay lengths in the problem, the tunneling amplitudes are
\begin{align}
\frac{t_\uparrow}{t_0}&\approx(1+i)\cos (2k_{SO}\ell+\pi/4)e^{-\kappa^{t}\ell}\,,\nonumber\\
\frac{t_\downarrow}{t_0}&\approx-(1-i)\cos (2k_{SO}\ell-\pi/4)e^{-\kappa^{t}\ell}\,,
\label{tun_small}
\end{align}
where the approximation neglects terms of order $1,~\kappa^{t}_1/\kappa^{t}\ll k_{SO}/\kappa^{t}$  and $t_0=(\pi L^2)^{1/4}k_{SO}\mathcal N\bar t_0/\kappa^{t}$ is the renormalized phenomenological constant which fixes the maximum tunneling. The functions in Eq.~(\ref{tun_small}) are plotted in Fig.~\ref{t_small} as a function of distance between the dot and topological section, $\ell$, for $L=0.01$, $k_{SO}=10$, and $\kappa^{t}=\sqrt{3}$; units of length are neglected as only the dimensionless product of lengths and wave vectors are relevant.
Notice that $\textrm{Re}[t_\uparrow]=\textrm{Im}[t_\uparrow]$ and $\textrm{Re}[t_\downarrow]=-\textrm{Im}[t_\uparrow]$, both of which decay exponentially with $\kappa^{t} \ell$ and oscillate with wavelength $\lambda_{SO}/2$,  which can be attributed to the relative factor of $e^{2 i k_{SO}x }$ between the dot and MF wave functions. Furthermore, because there is a difference in phase of $\pi/2$ between $t_\uparrow$ and $t_\downarrow$,  by changing the distance between the dot and MF, the tunneling can go from full polarization of one spin, either $t_\uparrow=0$ or $t_\downarrow=0$, to equal magnitude spin tunneling, $|t_\uparrow|=|t_\downarrow|$. 

\begin{figure}[tb]
\includegraphics[width=1\linewidth]{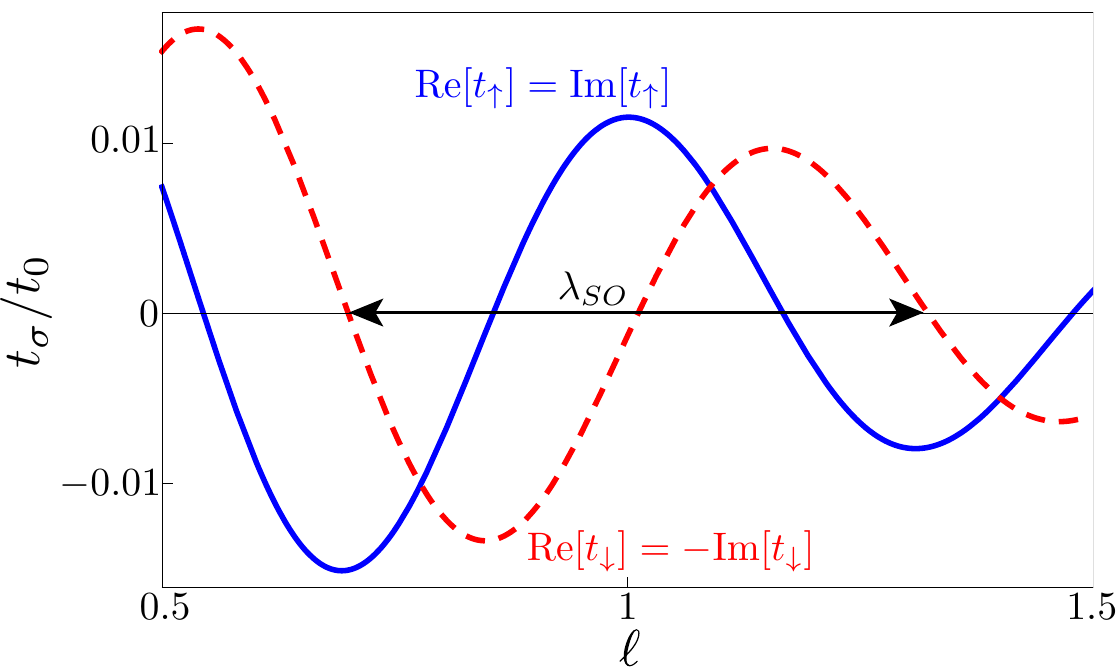}
\caption{The real and imaginary components of spin-dependent tunneling amplitudes $t_\uparrow$ (blue solid) and $t_\downarrow$ (red dashed) for a large dot, $L=1$, $k_{SO}=10$, and $\kappa^{t}=\sqrt{3}$. The oscillation wavelength is $1/\lambda_{SO}$ and the magnitude is smaller as compared with the small dot. Because the dot is of finite size,  we start with a separation $\ell=L/2=0.5$ between the center of the dot and the end of the topological section.}
\label{t_large}
\end{figure}

When the dot size is comparable to the spin orbit length, $k_{SO} L\gtrsim1$, there is no simple analytical formula for the tunneling coefficients. Upon comparing the dot and MF wave functions, there is a relative factor of $e^{i k_{SO}x}$ and we therefore expect the tunneling amplitudes to oscillate with the wavevector $k_{SO}$, which is half that of the small dot. Plotting $t_\sigma/t_0$, for $L=1$, $k_{SO}=10$, and $\kappa^{t}=\sqrt{3}$ in Fig.~\ref{t_large}, we see an exponential decrease as a function of $\ell$ and oscillatory behavior with period $\lambda_{SO}$ with the spin up and down components differing in phase by $\pi/2$. Because the wave function of the dot is extended, the overlap of dot and MF wave functions is reduced, as compared to the small dot case, so that the maximum magnitude of $t_\sigma/t_0$ is small; the magnitude of $t_\sigma$ can be increased by increasing $\bar t_0$ which roughly corresponds to decreasing the barrier between the topological end and the quantum dot in an experiment. 

\section{Effective Magnetic Field}
\label{readout}

In this section we extend our setup by considering two ends of identical topological superconductor sections, separated by a distance $d$, flanking opposite sides of a quantum dot [see Fig.~\ref{setup1}(b)], at a distance $\ell$ from the right MF. Because the MF wave functions are symmetric at the ends of either topological superconductor section, the overlap of the right tunneling amplitudes are given by Eq.~(\ref{tun_small}), $t_{\sigma r}=t_\sigma$, while the left tunneling amplitudes are analogously given by
\begin{align}
\frac{t_{\uparrow l}}{t_0}&\approx(1-i)\cos [2k_{SO}(d-\ell)+\pi/4]e^{-(d-\ell) \kappa^{t}}\,,\nonumber\\
\frac{t_{\downarrow l}}{t_0}&\approx-(1+i)\cos [2k_{SO}(d-\ell)-\pi/4]e^{-(d-\ell) \kappa^{t}}\,.
\end{align}
Here, we neglect any direct overlap between MFs in the wire\cite{pradaPRB12,rainisPRB13}
or via the bulk superconductor\cite{zyuzinPRL13}.
The corresponding tunneling Hamiltonian is written as \cite{hoffmanPRB16}
\begin{equation}
H_T=\sum_{\sigma,\lambda} t_{\sigma\lambda} d_\sigma^\dagger \gamma_\lambda +\textrm{H.c.}\,,
\end{equation}
where $\lambda=l,r$ specifies the left and right MF, respectively. Following Ref.~\onlinecite{hoffmanPRB16}, we find that a Schrieffer-Wolff transformation yields,\cite{schriefferPR66} to second order in tunneling,
\begin{equation}
\mathcal H_T=\sum_{i=0,\ldots,3} (B^{-}_i f f^\dagger+B^{+}_i f^\dagger f) S_i\,,
\label{tun_B}
\end{equation}
where $f=\gamma_r+i\gamma_l$ ($f^\dagger=\gamma_r-i\gamma_l$) is the nonlocal fermionic annihilation (creation) operator formed from the MFs and $S_i=\sum_{\sigma\sigma'} d^\dagger_\sigma\sigma^i_{\sigma\sigma'}d_{\sigma'}$ is the spin operator on the dot with $\sigma^0=\mathbb{1}_{2\times2}$. We remind the reader that the axis of quantization here is the \textit{dot} axis, along the applied magnetic field ($x$ axis), which is related to the \textit{wire} axis of quantization, along the spin orbit direction ($z$ axis), by a $\pi/2$ rotation around the $y$ axis.  According to Eq.~(\ref{tun_B}), a different effective magnetic field is exerted on the quantum dot when the fermionic state is occupied, $B^+_i$, or unoccupied, $B^-_i$, where (see Appendix~\ref{effH})
\begin{align}
B^{\pm}_0&=\frac{|t_{\uparrow\pm}|^2}{\epsilon_\uparrow\pm2\delta}+\frac{|t_{\downarrow\pm}|^2}{\epsilon_\downarrow\pm2\delta}\,,\nonumber\\
B^\pm_1&=-B^\pm_z=\textrm{Re}(t^*_{\uparrow\pm}t_{\downarrow\pm})\left(\frac{1}{\epsilon_{\uparrow}\pm2\delta}+\frac{1}{\epsilon_{\downarrow}\pm2\delta}\right)\,,\nonumber\\
B^\pm_2&=B^\pm_y=\textrm{Im}(t^*_{\uparrow\pm}t_{\downarrow\pm})\left(\frac{1}{\epsilon_{\uparrow}\pm2\delta}+\frac{1}{\epsilon_{\downarrow}\pm2\delta}\right)\,,\nonumber\\
B^\pm_3&=B^\pm_x=\frac{|t_{\uparrow\pm}|^2}{\epsilon_\uparrow\pm2\delta}-\frac{|t_{\downarrow\pm}|^2}{\epsilon_\downarrow\pm2\delta}\,,
\label{Bs}
\end{align}
\linebreak
$t_{\sigma\pm}=t_{\sigma l }\pm t_{\sigma r }/i$, and $\delta$ is the splitting due to the overlap of the MFs closest to the dot on the right, $\gamma_r$, and left, $\gamma_l$, topological section. Because of the direction of the quantization axis, $B_1^\pm$, $B_2^\pm$, and $B_3^\pm$ induces a spin splitting along minus the SOI axis ($-z$ axis), $y$ axis, and axis parallel to the applied magnetic field ($x$ axis), respectively. 

Let us consider the case when the ends of the topological sections are sufficiently far apart, $\delta=0$, for which 
\begin{widetext}
\begin{align}
\frac{B^\pm_1}{t_0}&=0\,,\,\,\,\,\frac{B^\pm_2}{t_0} \approx  \frac{\bar\epsilon_\uparrow+\bar\epsilon_\downarrow}{2\bar\epsilon_\uparrow\bar\epsilon_\downarrow}\left\{e^{-2\kappa^{t} \ell}\cos(4k_{SO}\ell)-e^{-2\kappa^{t} (d-\ell)}\cos[4k_{SO}(d-\ell)]\pm e^{-\kappa^{t} d}\sin\left[2k_{SO}\left(2\ell-d\right)\right]\right\}\,,\nonumber\\
\frac{B^\pm_3}{t_0}&\approx \frac{1}{\bar\epsilon_\uparrow} \left\{e^{-\kappa^{t}\ell}\cos[2k_{SO}\ell+\pi/4]\pm e^{\kappa^{t}(\ell-d) }\cos[2k_{SO}(\ell-d)-\pi/4]\right\}^2\nonumber\\
& -\frac{1}{\bar\epsilon_\downarrow} \left\{e^{-\kappa^{t}\ell}\cos[2k_{SO}\ell-\pi/4]\mp e^{\kappa^{t}(\ell-d) }\cos[2k_{SO}(\ell-d)+\pi/4]\right\}^2\,,
\label{B_full}
\end{align}
\end{widetext}
where the approximations assume $k_{SO}\gg \kappa^{t}_1,~\kappa^{t}$ with $\bar\epsilon_\sigma=\epsilon_\sigma/t_0\gg1$ for the perturbative Schrieffer-Wolff transformation to remain valid. Performing the following consecutive operations brings the system [Fig.~\ref{setup1}(b)] back to itself: mirror operation in the $yz$ plane, time reversal, and a $\pi$ rotation around the $y$ axis. These operations take $B_1^\pm\rightarrow -B_1^\pm$, while the other components are invariant. Therefore, $B_1^\pm$, \textit{i.e.}, the effective magnetic field along the spin-orbit axis, must be identically zero even for finite overlap of the MFs, $\delta\neq0$. As $B_2^+\neq B_2^-$ and $B_3^+\neq B_3^-$, these components are sensitive to the occupancy of the nonlocal fermion. Notice, however, when the quantum dot is far away from one end, e.g. $d\rightarrow\infty$, the effective magnetic field is insensitive to this quantity, $B_2^+= B_2^-$ and $B_3^+= B_3^-$, as one may expect.

When the ends of the topological sections are equidistant to the center of the dot, $d=2\ell$, the components of the effective magnetic field simplify to
\begin{align}
\frac{B^\pm_1}{t_0}&=\frac{B^\pm_2}{t_0}=0\,,\nonumber\\
\frac{B^\pm_3}{t_0}&\approx 8\frac{1\mp\sin(4k_{SO}\ell)]e^{-2\kappa^t\ell}}{(\epsilon_\uparrow-\epsilon_\downarrow)\pm(\epsilon_\uparrow+\epsilon_\downarrow)}\,.
\label{del}
\end{align}

\begin{figure}[t]
\includegraphics[width=.94\linewidth]{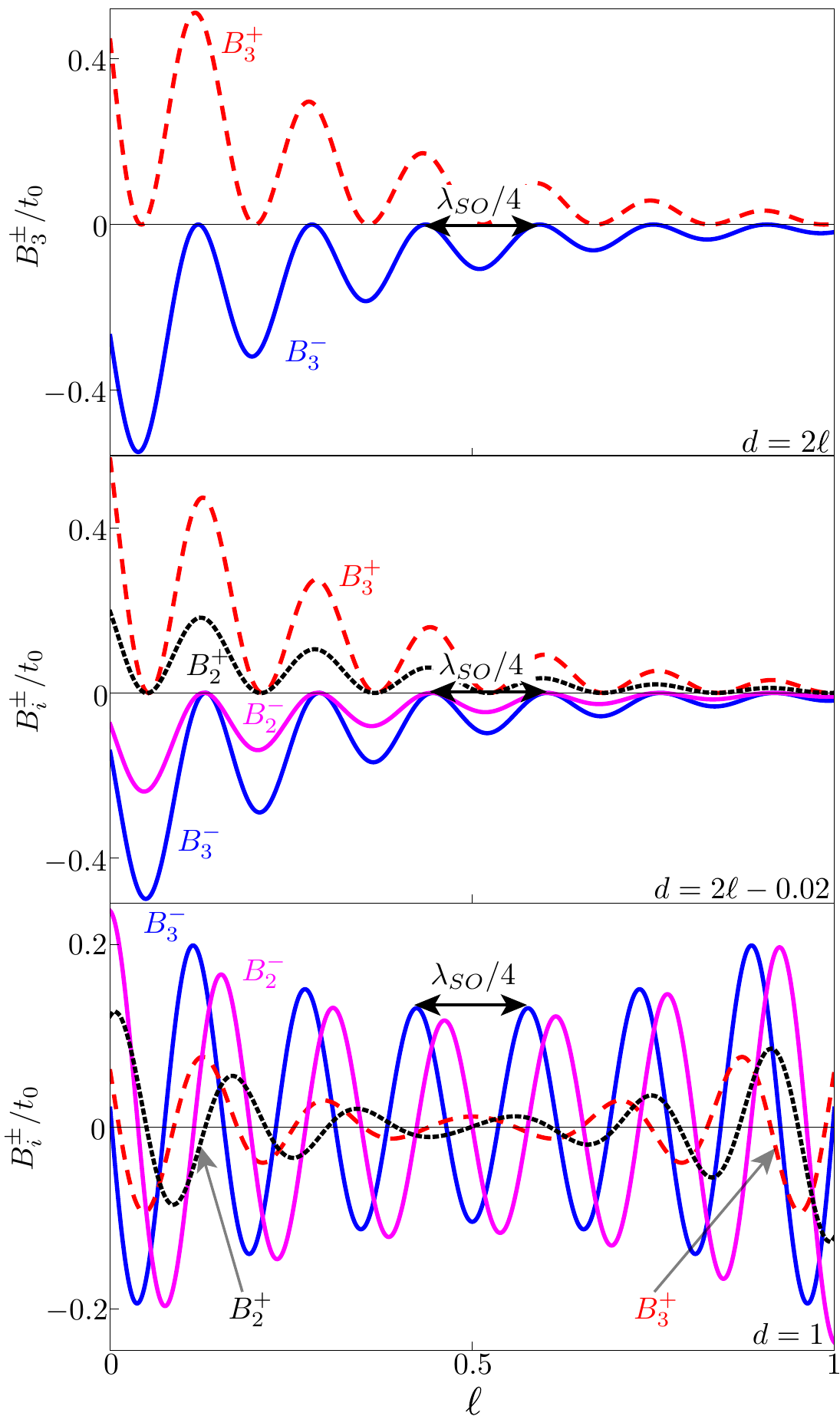}
\caption{Effective magnetic field, $B_i^\pm$, on a small dot situated  equidistant to two MFs, $d=2\ell$ (upper panel), slightly asymmetrically to the two MFs, $d=2\ell-0.02$ (middle panel), and  in the case when one fixes the distance between between MFs (lower panel), $d=1$ as a function of the distance, $\ell$, between the dot and topological sections for $L=0.01$, $k_{SO}=10$,  $\kappa^{t}=\sqrt{3}$, $\epsilon_\uparrow=9t_0$, and $\epsilon_\downarrow=11 t_0$. Upper panel: The components $B_3^\pm$ oscillate with period $\lambda_{SO}/4$ while the other components are zero. Middle panel: $B_2^\pm$ and $B_3^\pm$ oscillate with period $\lambda_{SO}/4$, $B^\pm_1=0$. Lower panel: $B^\pm_2$ is a asymmetric function of $\ell$ around $\ell=0.5$ while $B^\pm_3$ is a symmetric function of $\ell$ around $\ell=0.5$; both components oscillate with period $\lambda_{SO}/4$.}
\label{b_small3}
\end{figure}

When $d=2\ell$, one may show the system is invariant upon inversion centered at the dot followed by a $\pi$ rotation around the $x$ axis wherein $B_2^\pm\rightarrow-B_2^\pm$ and therefore must be zero. In Fig.~\ref{b_small3}  (upper panel), we plot the fields $B_3^\pm$ for $L=0.01$, $k_{SO}=10$, $\kappa^{t}=\sqrt{3}$, $\epsilon_\uparrow=9 t_0$ and $\epsilon_\downarrow=11 t_0$  as a function of $\ell$. As expected, we see that the component of the effective magnetic field along the axis of the applied magnetic field oscillates with period $\lambda_{SO}/4$ and changes according to the occupation of the nonlocal fermion.

If the center of the quantum dot is placed slightly asymmetrically, on the scale of the spin orbit length, between the ends of the topological sections, the effective magnetic field acquires a finite component along the $y$ axis. We plot $B_2^\pm$ and $B_3^\pm$ for this geometry in Fig.~\ref{b_small3} (middle panel) for $d=2\ell-0.2$, $L=0.01$, $k_{SO}=10$, $\kappa^{t}=\sqrt{3}$, $\epsilon_\uparrow=9 t_0$ and $\epsilon_\downarrow=11 t_0$  as a function of $\ell$. The component of the effect magnetic field along the $x$ axis is largely unchanged while the component along $y$ also oscillates with period $\lambda_{SO}/4$ but smaller amplitude. Furthermore, the local minima and maxima of $B_3^\pm$ and $B_2^\pm$ are shifted with respect to each other.

Fixing the distance between two topological sections, we plot the effective magnetic field as a function of distance between the dot and the right topological section in the lower panel of Fig.~\ref{b_small3} for the same values as the previous panels. Here, $B^\pm_2$ and $B^\pm_3$ both oscillate with period $\lambda_{SO}/4$ and $B^\pm_2=0$ at $\ell=1$ as expected. Again making use of inversion centered in the middle of the dot and a $\pi$ rotation around the $x$ axis, the distance between the dot and the left fermion ($d-\ell$) is exchanged with the right fermion ($\ell$) while the remainder of the geometry is invariant. Because of the transformations of the effective magnetic field under this symmetry, we expect $B^\pm_2$ ($B^\pm_3$) to be antisymmetric (symmetric) about $\ell=d/2$, which is readily observed [Fig.~\ref{b_small3} (lower panel)].

\begin{figure}
\includegraphics[width=1\linewidth]{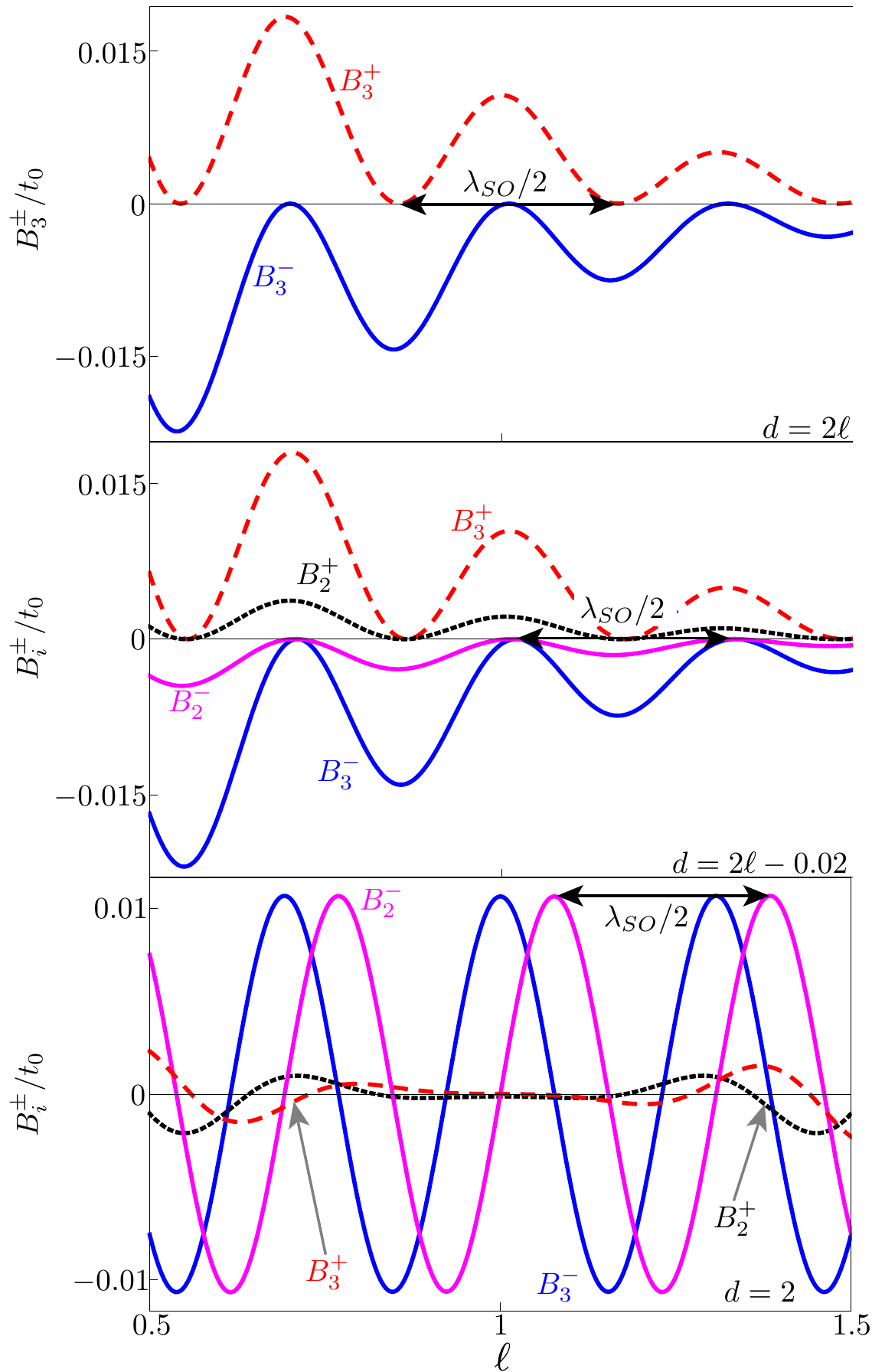}
\caption{
Effective magnetic field, $B_i^\pm$, on a large dot situated  equidistant to two MFs, $d=2\ell$ (upper panel),  slightly asymmetrically to the two MFs, $d=2\ell-0.02$ (middle panel), and in the case when one fixes the distance between between MFs (lower panel), $d=2$, as a function of the distance, $\ell$, between the dot and topological sections for $L=1$, $k_{SO}=10$,  $\kappa^{t}=\sqrt{3}$, and $\epsilon_\uparrow=\epsilon_\downarrow=t_0/10$. Upper panel: The components $B_3^\pm$ oscillate with period $\lambda_{SO}/2$ while the other components are zero. Middle panel: $B_2^\pm$ and $B_3^\pm$ oscillate with period $\lambda_{SO}/2$, $B^\pm_1=0$. Lower panel: $B^\pm_2$ is a asymmetric function of $\ell$ around $\ell=0.5$ while $B^\pm_3$ is a symmetric function of $\ell$ around $\ell=1$; both oscillate with period $\lambda_{SO}/2$.}
\label{b_large1b}
\end{figure}

\textit{Large dot.}--We plot the effective magnetic field induced on the large dot due to coupling to the MF states, for $L=1$, $k_{SO}=10$,  $\kappa^{t}=\sqrt{3}$, and $\epsilon_\uparrow=\epsilon_\downarrow=t_0/10$; because $t_\sigma/t_0\lesssim0.01\ll\epsilon_\sigma/t_0=0.1$ the perturbative expansion used to derive Eq.~(\ref{Bs}) remains valid. Analogous to the small dot, $B^\pm_i$ is plotted in the upper, middle, and lower panels of Fig.~\ref{b_large1b} when the left and right MFs are equidistant from the quantum dot ($d=2\ell$), when the quantum dot is placed slightly asymmetrically between the MFs ($d=2\ell-0.02$), and fixing the distance between between the MFs ($d=2$), respectively. Because the same symmetry arguments can be made, the large dot effective magnetic field is similar to the small dot with the important difference that the oscillations, as a function of $\ell$, are periodic with $\lambda_{SO}/2$ rather than $\lambda_{SO}/4$. In particular, when $d=2\ell$, only $B_3^+\neq B_3^-$ is finite; when $d=2\ell-0.02$, both $B_2^+\neq B_2^-$ and $B_3^+\neq B_3^-$; and fixing $d=2$, the $B_2^\pm$ is antisymmetric about $\ell=1$ and $B^\pm_3$ is symmetric about $\ell=1$. 

Experimentally, the bare tunneling must be less than the finite size energy spacing of the dot, $t_\sigma<\hbar\omega_0$, which is approximately $50~\mu$eV in typical experiments, corresponding to $t_05~m$eV. Therefore, the maximal value of $B^\pm_i$ that we expect to measure is $50~\mu$eV which is well above typical temperatures of 20~mK.

 \begin{figure}[t]
\includegraphics[width=9cm]{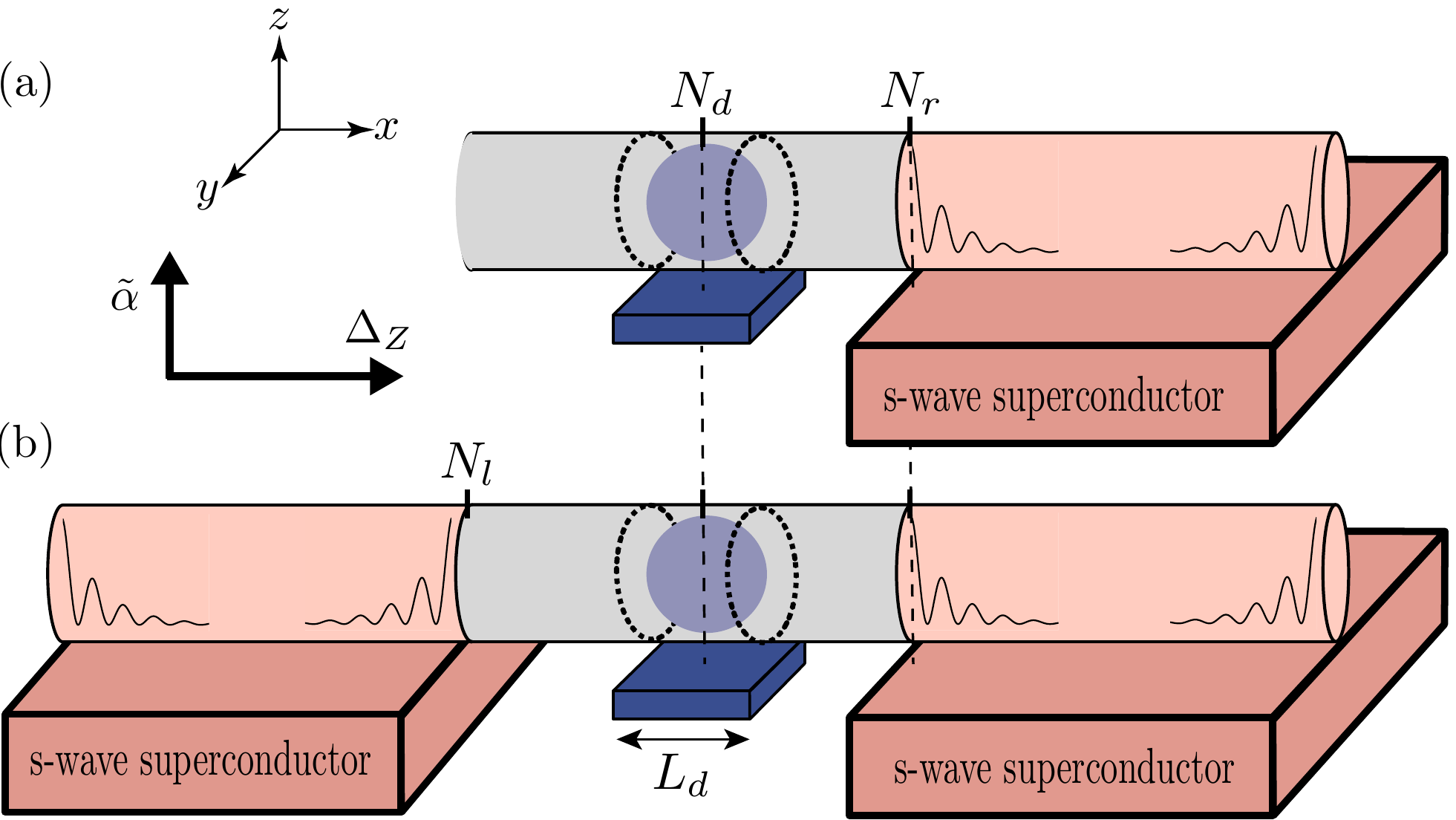}
\caption{System setup of the $N$-site tight-binding model. In the upper panel, the topological section (red), defined from site $N_r$ to the end of the chain, is realized due to proximity-induced superconductivity. The nontopological section (grey) is driven to the topologically trivial phase by depleting the wire. A quantum dot (blue) of size $L_d$ defined by gates is located at $N_d<N_r$. In the lower panel, the setup is the same with the addition of a second topological section, realized in the same way, from the beginning of the chain to $N_l<N_d$. In both setups, the magnetic field, inducing a Zeeman splitting $\Delta_Z$, is applied along the $x$ axis and the spin-orbit vector, with magnitude $\tilde\alpha$, is along the $z$ axis. }
\label{fig:system1}
\end{figure}

\section{Numerical Simulation}
\label{nums}
In this section, we numerically study effects resulting from the interplay between quantum dot weakly coupled to one or two wires hosting MFs in a physical system that one can experimentally engineer. As discussed above, it is difficult to determine the exact MF wave function when the chemical potential of the nontopological section is in the bandgap, so we have focused above on the situation when the nontopological section was created only by a slight detuning of the chemical potential from the SOI energy.  In contrast to that, the most viable way to terminate the topological section is to deplete a part of the quantum wire.  This scenario we can study numerically by using a tight-binding approach to calculate the spin-dependent tunneling between the dot and MF (Sec.~\ref{num_tuns}) and the spin polarization of the dot, which reflects the effective magnetic field, in the presence of two MF wires (Sec.~\ref{num_mags}). We confirm that our analytical results capture the main effects such as oscillations of tunneling amplitude as a function of distance.

\subsection{Spin-dependent tunneling }

\label{num_tuns}
We consider an $N$-site tight-binding Bogoliubov-de-Gennes Hamiltonian (see Fig. \ref{fig:system1}), analogous to our analytical model, 
\begin{align}
&H=\sum_{j=1}^{N-1}\Psi_{j+1}^\dag(-t-i\tilde{\alpha}\sigma_3)\eta_3\Psi_j+\textrm{H.c.} \label{TB_BdG} \\
&+\sum_{j=1}^N\Delta_Z\Psi_j^\dag\sigma_1\eta_3\Psi_j - \mu_j\Psi_j^\dag\eta_3\Psi_j+\Delta_{s,j}\Psi_j^\dag\sigma_2\eta_2\Psi_j, \nonumber
\end{align}
where we are in the Nambu basis $\Psi^\dagger_j=(\psi^\dagger_{j\uparrow}, \psi^\dagger_{j\downarrow}, \psi_{j,\uparrow}, \psi_{j,\downarrow})$ and the operator $\psi^\dagger_{j\sigma}$ creates a particle of spin $\sigma$ at site $j$. The hopping amplitude, $t = \hbar^2/(2ma^2)$, is set to 1 and taken as the energy unit and the SOI strength, $\tilde{\alpha}$, is fixed to $0.5$ for the remainder of the manuscript. The magnetic field (with Zeeman energy $\Delta_Z$) is aligned along the $x$ axis and is constant. The SOI vector points along $z$ axis, analogous to the previous section. The chemical potential, $\mu_j$, is $\mu_{t}$ for $j> N_r$ ({\it i.e.}, the topological section), $\mu_{n}$ for $j<N_d-L_d/2$ and $N_d+L_d/2<j\leq N_r$ (the nontopological section {\it excluding} the dot), and $\mu_d$ for $N_d-L_d/2\leq j\leq N_d+L_d/2$ which defines the quantum dot. The superconducting pairing is zero in the nontopological section, $\Delta_{s,j}=0$, for $j \leq N_r$ and $\Delta_s$ otherwise. 

\begin{figure}[b]
\includegraphics[width=8cm]{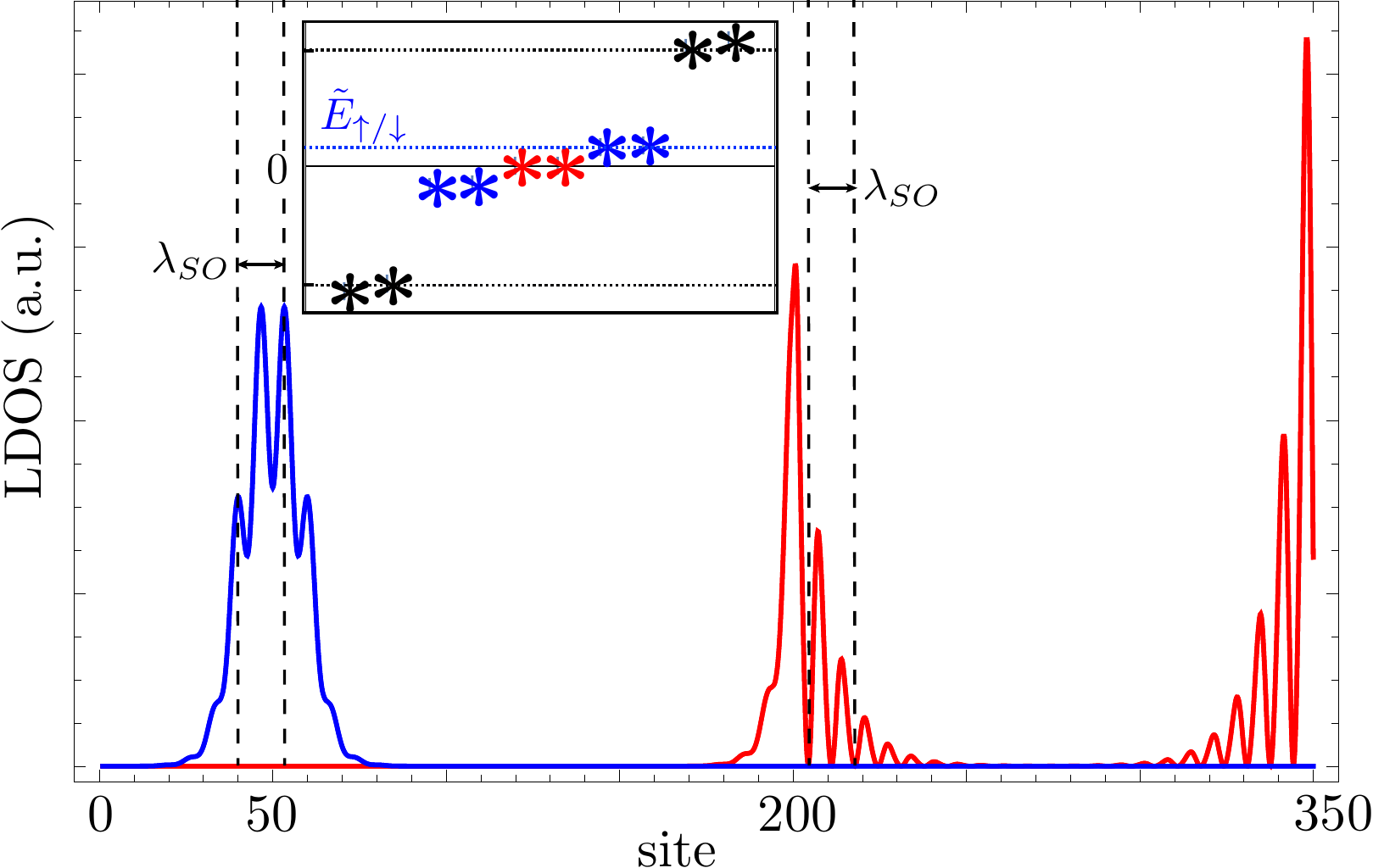}
\caption{Local density of states (LDOS) of the wire with one topological section [Fig.~\ref{fig:system1}(a)]. The lowest quantum dot level (blue) is centered at $N_d=50$ and of the size $L_d=31$ while the MF wave functions (red) are peaked roughly at the beginning ($N_r=200$) and at the end ($N=350$) of the topological section. Here, $\Delta_s=0.06$, $\Delta_Z=0.12$, $\mu_{n}=2.27$, $\mu_{t}=2$ and $\mu_d=2.245$. Inset: Energy spectrum indicates two MF levels (red) at zero energy, two quantum dot levels (blue) with energies at $\tilde E_\uparrow\approx \tilde E_\downarrow$, and the next lowest energy levels of the dot (black).}
\label{Fig_spin}
\end{figure}

We take a wire of length $N=350$ lattice sites. Sites beyond $N_r=200$ are driven into the topological phase by taking $\Delta_s=0.06$, $\Delta_Z=0.12$, $\mu_{t}=2$. In the nontopological section of the wire the chemical potential is $\mu_{n}=2.27$ except on the dot, which is of size $L_d=31$, where $\mu_d=2.245$. Upon diagonalizing the Hamiltonian, we find the position-dependent wave functions and corresponding eigenenergies which are plotted in Fig.~\ref{Fig_spin}. The dot wave function, centered at $N_d=50$, is Gaussian-like with oscillations at $\lambda_{SO}/2=\pi/\tilde\alpha\approx2\pi$ which correspond to roughly half the spin-orbit length. There are two MFs, one at each end of the topological section. Within the topological section, the MF oscillates with period $\lambda_{SO}/2$. On the left side, the MF wave function `leaks' into the normal section and has oscillations, also given by the SOI and a smaller relative amplitude proportional to the magnetic field as expected from the analytics [see Eq.~(\ref{NT_IG})]. It is precisely this leakage and oscillations that results in a position dependent tunneling, and subsequent magnetic field, that we discuss below. In the spectrum, there are two zero energy modes in the center of the plot (see the insert in Fig.~\ref{Fig_spin}) corresponding to the MFs at the ends of the wire. The next two lowest lying energies above zero energy are the spin up and down, nearly degenerate states on the dot. Because the dot size is much larger than the spin orbit length, the magnetic field on the dot is exponentially suppressed and the Zeeman splitting on the dot is nearly zero.\cite{trifPRB08}

\begin{figure}[t]
\includegraphics[width=8cm]{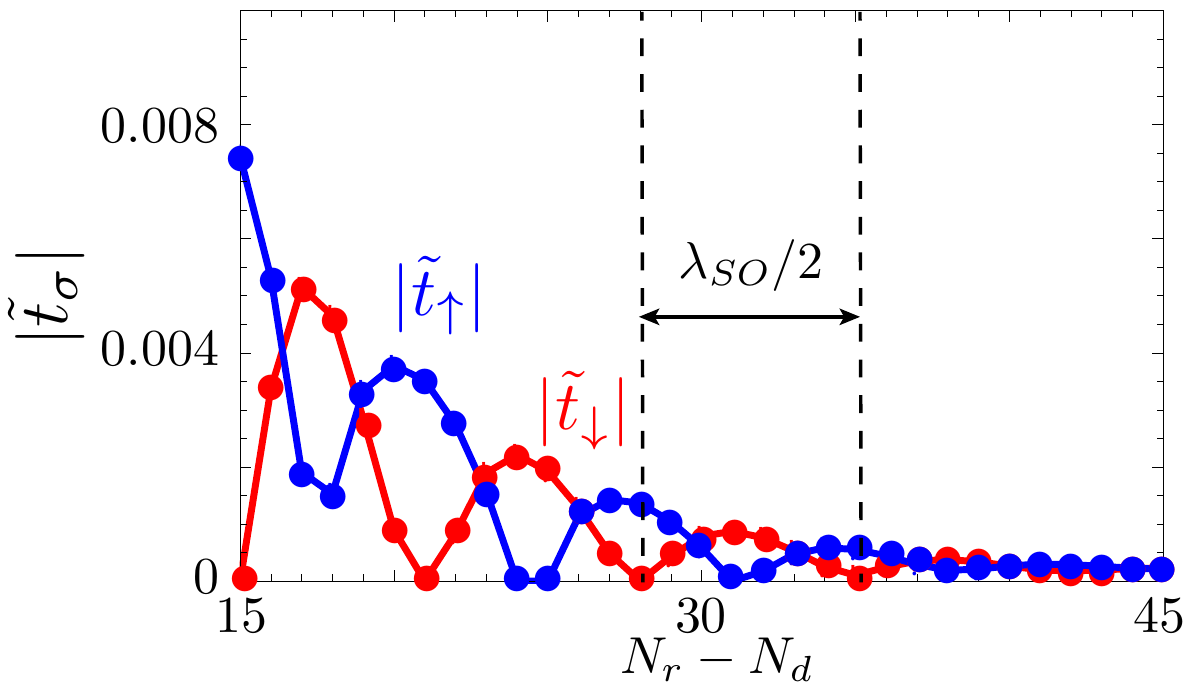}
\caption{Tunneling amplitudes $|t_\uparrow|$ and $|t_\downarrow|$ between the quantum dot and MF [Fig.~\ref{fig:system1}(a)] as a function of the distance between the position of the dot and the end of the topological section of the wire. All parameters are the same as in Fig. \ref{Fig_spin}.
}
\label{fig:tspin1}
\end{figure}

To extract the spin-dependent tunneling amplitude $\tilde t_\sigma$ with $\sigma=\uparrow,\downarrow$, we model the dot-MF system as two weakly coupled levels. Here, because we do not have access to the quantization axis of the dot, $\sigma$ labels the two dot levels. If MFs leaks into the dot, the dot level is shifted  from $\tilde\epsilon_\sigma$  to $\tilde E_\sigma=\sqrt{\tilde\epsilon_\sigma^2+2\tilde t_\sigma^2}$ where $\tilde{\epsilon}_{\sigma}$ is the energy of the spin $\sigma$ level when the dot is far from the MF. Thus, we can extract the spin-dependent coupling $|\tilde t_{\sigma}|=\sqrt{\left(\tilde E^2_{\sigma}-\tilde\epsilon_{\sigma}^2\right)/2}$, which we plot as a function of distance between the dot and MF in Fig.~\ref{fig:tspin1}. As expected, the tunneling amplitude $|\tilde t_\sigma|$ decreases exponentially and oscillates with period $\lambda_{SO}/2$ as the distance between the dot and MF increases. Furthermore, the tunneling amplitudes are offset from each other by a phase $\pi/2$, in agreement with the analytics. Because we are only probing the change in energy of the dot level, we can only determine the magnitude of the spin-depedent tunneling; we expect that $\tilde t_\sigma$ to oscillate with periodicity $\lambda_{SO}$. 

\subsection{Effective magnetic field}
\label{num_mags}

Next, we extend our model [see Eq.~(\ref{TB_BdG})] and add an additional topological section to the left of the dot  [see Fig.~\ref{fig:system1}(b)].
 The site-dependent parameters are redefined as follows: The chemical potential is $\mu_{t}$ for $j\leq N_l$  and $j> N_r$ where $N_l$ now defines the end of the second topological section, $\mu_{n}$ for $N_l<j<N_d-L_d/2$ and $N_d+L_d/2<j\leq N_r$, and $\mu_d$ for $N_d-L_d/2\leq j\leq N_d+L_d/2$. The superconducting pairing is zero, $\Delta_{s,j}=0$, for $N_l<j\leq N_r$ and $\Delta_s$ otherwise. 

\begin{figure}[t]
\includegraphics[width=8cm]{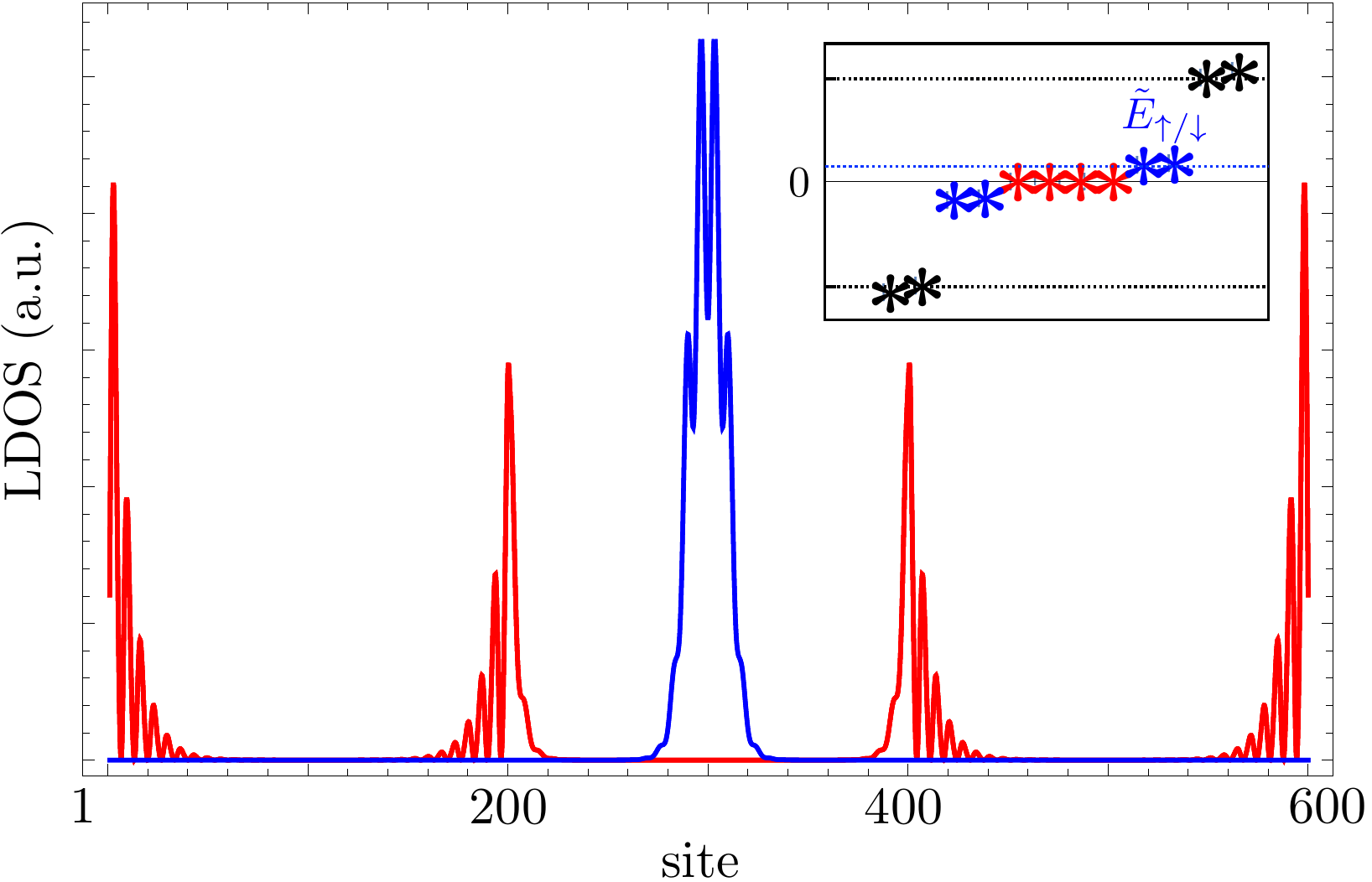}
\caption{Local density of states of a chain with two identical topological sections [Fig.~\ref{fig:system1}(b)] between $N=1$ and $N_l=200$ and between  $N_r=400$ and the end of the chain ($N=600$). The quantum dot wave function (blue) is centered around the dot position $N_d=300$ and there are four MFs  at the interfaces of the topological and nontopological sections. Inset: Energy spectrum indicates  four zero energy states corresponding to the MFs (red), the lowest energy dot levels (blue), and second lowest dot levels (black). The system parameters are the same as in Fig. \ref{Fig_spin}. }
\label{fig:eigenB}
\end{figure}

We now take $N=600$, $N_d=300$ with $N_r$ and $N_l$ free to vary. All other parameters are left unchanged. The sites with $j\leq N_l$ and $j>  N_r$ are in the topological regime. Plotting the wave functions (see Fig.~\ref{fig:eigenB}), we see, accordingly, that there are indeed four MF states at the four interfaces of the topological with nontopological sections, all of which sit at zero energy. The characteristics of MF and dot level wave functions (delay lengths and period of oscillations) are the same  as in the previous subsection.

\begin{figure}[t]
\includegraphics[width=8cm]{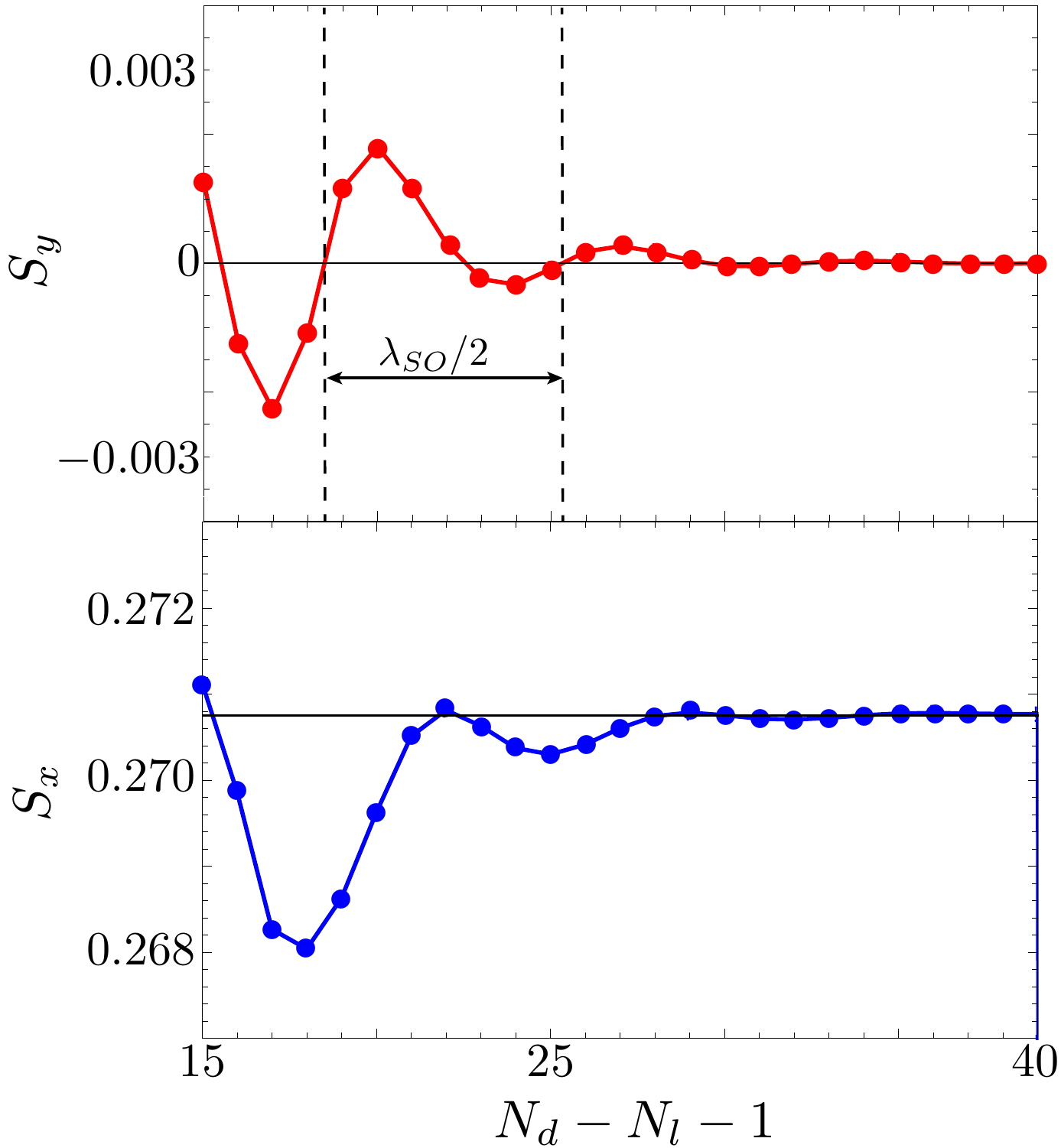}
\caption{The spin components of the lowest quantum dot level along the $y$ and $x$ axes, respectively, as a function of the distance between the dot and end of the left and right topological sections, which are kept equidistant to the dot up to one lattice constant, $N_d-N_l-1=N_r-N_d-1$. Both components oscillates with period $\lambda_{SO}/2$, and depend exponentially on the distance between  the dot and the MFs.
The spin projection on the $y$ axis goes to zero when the dot is far from MFs while the $x$ component saturates at the value determined by the external magnetic field (black solid line around $\approx0.271$). We note that the component $S_z$ is always zero due to the symmetry of the problem.}
\label{fig:SxSz}
\end{figure}

To extract the effective magnetic field on the dot, we calculate the spin of the dot by summing the expectation of the spin operator, $\hat S_{x}=\sigma_1\eta_3$, $\hat S_{y}=\sigma_2$, and  $\hat S_{z}=\sigma_3\eta_3$, at all sites where the dot level has finite weight.  In Fig.~\ref{fig:SxSz}, we present the spin on the dot, $S_{i}=\sum_j\tilde Y^\dagger_j \hat S_i \tilde Y_j$ (measured in units of $\hbar/2$) with $\tilde Y_j$ the dot wave function at site $j$, as a function of the distance between the dot and  MFs. 
Analogous to the previous section, we have considered nearly symmetrically placed topological sections so that the MF on the left and right are equidistant to the dot up to one lattice constant, {\it i.e.}, $N_{d}-N_l-1=N_r-N_{d}-1$. Similar to the analytic results, we see oscillations in spin on the dot with period $\lambda_{SO}/2$ along the $x$ and $y$ axes while the spin along the SOI axis is exactly zero. The offset of $S_x$ in Fig. \ref{fig:SxSz} is the result of a residual magnetic field coming from the applied external Zeeman field along the $x$ direction (see Appendix \ref{num_residual} for details). We note that in Fig. \ref{fig:SxSz}, because $L_d$ is odd and $N_r-N_l$ is even, the dot is closer to the left topological section than to the right topological section by one lattice constant. As a result, $S_y \neq 0$, which is consistent with our analytical predictions. If the dot is placed equidistantly between the two topological sections, $S_y$ is zero.

In contrast to the analytic results, there are two important differences in the tight-binding calculation: (1) we are unable to account for many-body interactions and therefore cannot differentiate between a filled and unfilled nonlocal fermion nor can we include a finite Coulomb interaction on the dot; and (2) the difference in physical realizations of the topological-nontoplogical junctions. Despite these differences between the models, we find a striking similarity in the spin-dependent tunneling and effective magnetic field. We attribute this to the equality of the symmetries in the analytic and numerical models. Therefore, we expect any MF-quantum dot system that obeys such symmetries, regardless of how the topological and nontopological regimes are realized, to display similar behavior of the spin-dependent tunneling and effective magnetic field.

\section{Conclusions and Outlook}

We have shown analytically and numerically that the tunneling amplitudes between MFs and a nearby quantum dot are spin-dependent and also depends on the distance between the dot and topological section hosting MFs. Generally, the spin up and down tunneling amplitudes are oscillating on the scale of the SOI length. In particular, depending on this distance, the tunnel amplitudes can be made to be completely spin-polarized. Analogously, the effective magnetic field induced on a quantum dot by two MFs depends on the distance between topological sections and quantum dot and, unlike the tunneling, on the occupancy of the nonlocal fermion formed from the MFs.

As a result, any phenomenological Hamiltonian between MFs and quantum dots must include a spin dependence in the tunneling in order to be applied to quantum wires. When the SOI length is large and the boundary between topological and nontopological sections or quantum dot are mobile, one could use the relative positions of the two as a way to fine tune the spin dependence of the tunneling. Alternatively, if the relative positions are fixed or the SOI length is smaller than the experimental precision, the spin dependence cannot be adjusted and thus may be a source of error. This is especially problematic when combining braiding and readout of MF qubits using quantum dots.~\cite{hoffmanPRB16} That is, after a braiding operation, the distance between the MF and quantum dot must be brought back to a precise position. If not, the qubit readout must be recalibrated. 

In lieu of a quantum wire, one could use a magnetic atomic chain deposited on the surface of a superconductor which has been theoretically\cite{nadj-pergePRB13,klinovajaPRL13,brauneckerPRL13,vazifehPRL13} and experimentally\cite{nadj-pergeSCI14,pawlakNPJ16} shown to support MF end states. The local helical magnetic field of the helical chain is equivalent to the SOI and homogeneous magnetic field. An auxiliary two level atom coupled to the ends of two such chains, analogous to the dot in our quantum wire setup, could be used to probe these MFs. We foresee two mechanisms by which the auxiliary atom can couple to the chain: the overlap of wave function of the orbital levels in the auxiliary atom with either the hybridized conduction bands in the atomic chain or with the bulk quasiparticles in the superconductor. When there is a direct tunneling between the orbital levels of the chain and the dot, we expect only the magnitude of the tunneling between MFs and levels in the auxiliary atom to vary as a function of the distance between the two because there is no analog of the SOI or magnetic field outside the chain. If there is a SOI in the superconductor, the spin-dependence of the tunneling could depend on the distance between the chain and auxiliary atoms, analogous to the role of the SOI in the quantum wire. We also note that instead of a quantum dot levels,  alternatively, one can also use finite-energy bound states inside the superconducting gap, for example, occurring due to change in the direction of the SOI vector.~\cite{klinovajaEPJB15} Again, we expect that the overlap between such bound states and MFs decays exponentially with the distance as well as oscillates on the scale set by the SOI length.

{\it Acknowledgements.} 
We acknowledge helpful discussions with Constantin Schrade, Pawel Szumniak, and Marcel Serina.  This work was supported by the Swiss NSF and  NCCR QSIT. 

\begin{appendix}

\begin{widetext}

\section{Majorana fermion wave functions}\label{wave_analytic}

\subsection{Small deviations of the chemical potential from the SOI energy}
In this Appendix we derive the MF wave function given in the main text [see Eq.~(\ref{maj_wav})].  The general wave functions on the left (nontopological section) and right (topological section) side are written as
\begin{align}
\Phi^t&=A^t\Phi_1^t+B^t\Phi_2^t+C^t\Phi_3^t+D^t\Phi_4^t\,,~~
\Phi^n=A^n\Phi_1^n+B^n\Phi_2^n+C^n\Phi_3^n+D^n\Phi_4^n\,,
\label{MF_bounds}
\end{align}
respectively, where the coefficients must be real if the solutions are MFs. To satisfy continuity, the coefficients must satisfy the equations
\begin{align}
A^t+C^t=-\cos\varphi_n A^n - \sin\varphi_n B^n + C^n\,,~~
A^t+C^t=\cos\varphi_n A^n+\sin\varphi_n B^n+C^n\,,\nonumber
\end{align}
which we have obtained by taking the imaginary part of the first component and the real part of the second component, respectively. This implies that $\cos\varphi_n A^n+\sin\varphi_n B^n=0$ and $C^n=A^t+C^t$. Using the former condition, differentiability of the solutions requires
\begin{align}
-(\kappa^n_1-\kappa^n_2)\cos\varphi_n A^n+\kappa^n C^n+2k_{SO}D^n = -\kappa^t_1A^t-\kappa^t C^t+2k_{SO}D^t\,,\nonumber\\
(\kappa^n_1-\kappa^n_2)\cos\varphi_n A^n+\kappa^nC^n+2k_{SO}D^n= -\kappa^t_1A^t-\kappa^t C^t+2k_{SO}D^t\,.\nonumber\\
\label{diff1}
\end{align}
Therefore, because $\kappa^n_1\neq\kappa^n_2$ and $\varphi_n\neq0$, $A^n=B^n=0$. Continuity further implies, after taking the real and imaginary parts of the first and second components, respectively,
\begin{align}
D^n=B^t+D^t\,,~~D^n=-B^t+D^t\,,
\end{align}
so that $B^t=0$ and $D^n=D^t$. With Eq.~(\ref{diff1}), one may show that $C^t=-A^t(\kappa^n+\kappa^t_1)/(\kappa^n+\kappa^t)$ and $C^n=A^t(\kappa^t-\kappa^t_1)/(\kappa^n+\kappa^t)$. Finally, invoking differentiability, one finds
\eq{
-2k_{SO} C^n+D^t\kappa^n=-\kappa_{SO}C^t-\kappa^t D^t\,,
}
and $D^t=2k_{SO}(C^n-C^t)/(\kappa^n+\kappa^t)=2k_{SO}/(\kappa^n+\kappa^t)$. Thus, we recover Eq.~(\ref{maj_wav}) in the main part.

\subsection{ Chemical potential is in the bandgap (depletion)}
\label{LCP}

When the nontopological section is characterized by the chemical potential being inside the bandgap such that this section is depleted, the wave functions are different than ones found above. In order to find these wave functions, we assume that $\mu_l$ is much larger than he SOI energy, the superconducting gap, and the magnetic field (the last too are put to zero in the nontopological section). In this case, we find that the eigenstates of the Hamiltonian are
\begin{equation}
\Psi^\pm_k=\psi^\pm e^{i k x}\,,\,\,\,\,\mathcal X^\pm_k=\chi^\pm e^{i k x}\,,
\label{ntsol}
\end{equation}
with energies $(k^2/2m-\mu_l)\pm\alpha k$ and $-[(k^2/2m-\mu_l)\pm\alpha k]$, respectively,
where $(\psi^+)^T=(1,\,0,\,0,\,0)$, $(\psi^-)^T=(0,\,1,\,0,\,0)$, $(\chi^+)^T=(0,\,0,\,1,\,0)$, and $(\chi^-)^T=(0,\,0,\,0,\,1)$. The zero energy solutions of Eq.~(\ref{ntsol}) require $k\equiv k^\mp=\mp k_{SO}-i \kappa_F$, where $\kappa_F=\sqrt{2 m \mu_l}/\hbar$ and we have chosen solutions that vanish as $x\rightarrow-\infty$.  It is easiest to match the topological section by finding linear superpositions that are MFs,\begin{equation}
\Phi_1^n=-i \Psi^++\Psi^-+i\mathcal X^++\mathcal X^-,\,\,\Phi^n_2=\Psi^+-i\Psi^-+\mathcal X^++i\mathcal X^-,\,\, \Phi^n_3=i \Psi^++\Psi^--i\mathcal X^++\mathcal X^-,\,\,\Phi^n_4=\Psi^-+i\Psi^-+\mathcal X^+-i\mathcal X^-,
\end{equation}
where $\Psi^\pm=\Psi^\pm_{\mp k}$ and $\mathcal X^\pm=\mathcal X^\pm_{\mp k }$.

Because these are now of the form of the MFs in the nontopological section, it is straightforward to find conditions for continuity of the MFs at the boundary which are $A^n=0$, $C^n=C^t+A^t$, $B^t=B^n$, and $D^t=D^n$, where the coefficients have been defined analogous to Eq.~(\ref{MF_bounds}). Upon solving the conditions for differentiability, we find $B^t=0$, 
\begin{align}
C^t=-A^t\frac{3 k_{SO}^2+\kappa_F^2+\kappa_F\kappa^t+\kappa_F\kappa_1^t+\kappa^t\kappa^t_1}{9k_{SO}^2+(\kappa_F+\kappa^t)^2}\,,\,\,\,\,D^t=A^t\frac{k_{SO}(\kappa_F-\kappa^t+3\kappa^t_1)}{9k_{SO}^2+(\kappa_F+\kappa^t)^2}\,.
\end{align}
Therefore, the MF wave function is given by
\begin{align}
\Phi^t&=A^t\left(\Phi_1^t-\frac{3 k_{SO}^2+\kappa_F^2+\kappa_F\kappa^t+\kappa_F\kappa_1^t+\kappa^t\kappa^t_1}{9k_{SO}^2+(\kappa_F+\kappa^t)^2}\Phi^t_3+\frac{k_{SO}(\kappa_F-\kappa^t+3\kappa^t_1)}{9k_{SO}^2+(\kappa_F+\kappa^t)^2}\Phi^t_4\right)\,,\\
\Phi^n&=A^t\left(\frac{6 k_{SO}^2+(\kappa^t)^2+\kappa_F\kappa^t-\kappa_F\kappa_1^t-\kappa^t\kappa_1^t}{9k_{SO}^2+(\kappa_F+\kappa^t)^2}\Phi^n_3+\frac{k_{SO}(\kappa_F-\kappa^t+3\kappa^t_1)}{9k_{SO}^2+(\kappa_F+\kappa^t)^2}\Phi^n_4\right)\,.
\end{align}

Adding a magnetic field perturbatively, to first order in the Zeeman energy $\Delta_Z$ the energies are unchanged, while the eigenvectors are transformed as
\begin{equation}
\Psi^\pm_k\rightarrow\tilde\psi^\pm_k=\Psi^\pm_k\pm\Delta_Z\Psi_k^\mp/2\alpha k\,,\,\,\,\,\mathcal X_k^\pm\rightarrow\tilde\chi^\pm_k=\mathcal X_k^\pm\pm\Delta_Z\mathcal X_k^\mp/2\alpha k\,.
\end{equation}
The zero energy solutions are thus $\tilde\psi^\pm=\tilde\psi^\pm_{\mp k}$ and $\tilde\chi^\pm=\tilde\chi^\pm_{\mp k}$. It is convenient to define
\begin{equation}
\tilde\Psi^\pm=\tilde\psi^\pm\mp\Delta_Z\tilde\psi^\mp/k^\mp\,,\,\,\,\,\tilde{\mathcal X}^\pm=\tilde\chi^\pm\pm\Delta_Z\tilde\chi^\mp/k^\pm\,,
\end{equation}
so that, to leading order in the Zeeman splitting, $\tilde\psi^\pm|_{x=\ell}=\psi^\pm$ and $\tilde\chi^\pm|_{x=\ell}=\chi^\pm$. We find the zero energy MFs  analogously,
\begin{equation}
\tilde\Phi_1^n=-i \tilde\Psi^++\tilde\Psi^-+i\tilde{\mathcal X}^++\tilde{\mathcal X}^-\,,\,\,\,\,\Phi^n_2=\tilde\Psi^+-i\tilde\Psi^-+\tilde{\mathcal X}^++i\tilde{\mathcal X}^-\,,\,\,\,\, \Phi^n_3=i \tilde\Psi^++\tilde\Psi^--i\tilde{\mathcal X}^++\tilde{\mathcal X}^-\,,\,\,\,\,\Phi^n_4=\tilde\Psi^-+i\tilde\Psi^-+\tilde{\mathcal X}^+-i\tilde{\mathcal X}^-\,,
\end{equation}
or in a more suggestive form
\begin{align}
\bar\Phi^{n}_1&=\left( \begin{array}{c}-ie^{-i k_{SO} (x-\ell)}+i\mathcal{S}^+\\e^{i k_{SO} (x-\ell)}+\mathcal{S}^-\\ i e^{i k_{SO} (x-\ell)}-i \mathcal{S}^-\\e^{-i k_{SO} (x-\ell)}+\mathcal{S}^+\end{array} \right) e^{\kappa_F (x-\ell)}\,,
\bar\Phi^{n}_2=\left( \begin{array}{c}e^{-i k_{SO} (x-\ell)}+\mathcal{S}^+\\-i e^{i k_{SO} (x-\ell)}+i \mathcal{S}^-\\ e^{i k_{SO} (x-\ell)}+\mathcal{S}^-\\i e^{-i k_{SO} (x-\ell)}-i\mathcal{S}^+\end{array} \right) e^{\kappa_F (x-\ell)}\,,\nonumber\\
\bar\Phi^{n}_3&=\left( \begin{array}{c}ie^{-i k_{SO} (x-\ell)}+i\mathcal{S}^+\\e^{i k_{SO} (x-\ell)}-\mathcal{S}^-\\ -i e^{i k_{SO} (x-\ell)}- i \mathcal{S}^-\\e^{-i k_{SO} (x-\ell)}-\mathcal{S}^+\end{array} \right) e^{\kappa_F (x-\ell)}\,,
\bar\Phi^{n}_4=\left( \begin{array}{c}e^{-i k_{SO} (x-\ell)}-\mathcal{S}^+\\ie^{i k_{SO} (x-\ell)}+i\mathcal{S}^-\\ e^{i k_{SO} (x-\ell)}- \mathcal{S}^-\\-ie^{-i k_{SO} (x-\ell)}-i\mathcal{S}^+\end{array} \right) e^{\kappa_F (x-\ell)}\,,
\label{NT_IG}
\end{align}
where $\mathcal{S}^\pm=\Delta_Z \sin[(x-\ell)k_{SO}]/\alpha k^\pm$. Because the MFs now have contributions from both left and right moving branches in the nontopological section of the wire, $|\bar\Phi_i|^2$ oscillates with periodicity proportional to $k_{SO}$ and amplitude $\Delta_Z$. Therefore, Eq.~(\ref{NT_IG}) suggests, in contrast to the small chemical potential, the probability of the MF wave function satisfying the boundary conditions also oscillates in the nontopological section. Although we find continuous and differentiable solutions when the magnetic field is zero, the condition of differentiability breaks down for finite magnetic field (Appendix \ref{LCP}). We focus on an analytic study in the regime of a small chemical potential and study the large chemical potential regime in Sec.~\ref{nums}, where we use a numerical tight-binding approach.

\section{Effective coupling between dot and MFs}

\label{effH}
In this Appendix we calculate the effective exchange Hamiltonian between the quantum dot levels and MFs by generalizing the work done in Ref.~\onlinecite{hoffmanPRB16}, which calculated the an effective exchange Hamiltonian for spin-independent tunneling amplitudes, for spin-dependent ones, $\tilde t_{\lambda\sigma}$. Following that reference, we take consider a system of two finite size TSC where $\tilde\gamma_\lambda$ and $\tilde\gamma_\lambda'$ are the MFs in the left and right ends, respectively, of wire $\lambda$ where the total Hamiltonian describing this system is defined by
\begin{align}
&H=\tilde H_M+H_D+ \tilde H_T\,,\nonumber\\
&\tilde H_M=i\sum_\lambda\tilde\delta_\lambda\tilde\gamma_{\lambda}'\tilde\gamma_{\lambda}\,,\nonumber\\
&H_D=\sum_\sigma\epsilon_\sigma d_\sigma^\dagger d_\sigma+Un_\sigma n_{\bar\sigma}/2\,,\nonumber\\
& \tilde H_T=\sum_{\sigma,\lambda}d_\sigma^\dagger(i\tilde t_{\sigma\lambda}'\tilde\gamma_{\lambda}'+\tilde t_{\sigma\lambda}\tilde\gamma_{\lambda})+(\tilde t^*_{\sigma\lambda}\tilde\gamma_{\lambda}-it'^*_{\sigma\lambda}\tilde\gamma_{\lambda}')d_\sigma \,.
\end{align} 
Here, $\tilde\delta_\lambda$ is the splitting of the MFs in TSC $\lambda$,  $U$ is the Coulomb repulsion on the dot, and $\tilde t'_{\sigma\lambda}$ ($\tilde t_{\sigma\lambda}$) is the matrix element for an electron with spin $\sigma$ on the dot tunneling to the MF in the  left (right) end of the $\lambda$th TSC. We rewrite the Majorana fermions as $ \tilde f_\lambda=(\tilde \gamma_{\lambda}'+i\tilde \gamma_{\lambda})/2$ so that $\tilde f^\dagger_\lambda \tilde f_\lambda=(1+i\tilde \gamma'_{\lambda}\tilde \gamma_{\lambda})/2$ and $i\tilde\delta_\lambda\tilde \gamma_{\lambda}'\tilde \gamma_{\lambda}=\tilde\delta_\lambda(2\tilde f_\lambda^\dagger \tilde f_\lambda -1)$. The logical values of the MF qubit are written in terms of the parity of the left and right wires. Using $\tilde \gamma_{\lambda}'=\tilde f_\lambda+\tilde f^\dagger_\lambda$ and $\tilde \gamma_{\lambda}=(\tilde f_\lambda-\tilde f^\dagger_\lambda)/i$, the tunneling Hamiltonian is transformed into
\begin{align}
\tilde \tilde H_T&=\sum_{\sigma\lambda}d_\sigma^\dagger[i\tilde t_{\sigma\lambda}(\tilde f_\lambda+\tilde f_\lambda^\dagger)-it'_{\sigma\lambda}(\tilde f_\lambda-\tilde f_\lambda^\dagger)]+[-it'^*_{\sigma\lambda}(\tilde f_\lambda-\tilde f_\lambda^\dagger)-i\tilde t^*_{\sigma\lambda}(\tilde f_\lambda+\tilde f_\lambda^\dagger)]d_\sigma\nonumber\\
&=\sum_{\sigma\lambda}i(\tilde t_{\sigma\lambda}'^*-\tilde t_{\lambda}^*)\tilde f^\dagger_\lambda d_\sigma - i(\tilde t_{\sigma\lambda}'^*+\tilde t_{\sigma\lambda}^*)\tilde f_{\lambda} d_\sigma+i(\tilde t_{\sigma\lambda}-\tilde t_{\lambda}') d_\sigma^\dagger \tilde f_\lambda + i(\tilde t_{\sigma\lambda}'+\tilde t_{\sigma\lambda}) d_\sigma^\dagger \tilde f_{\lambda}^\dagger\nonumber\\
&=\sum_{\sigma\lambda}i\tilde t_{\sigma\lambda -}^*\tilde f^\dagger_\lambda d_\sigma - i\tilde t_{\sigma\lambda +}^*\tilde f_{\lambda} d_\sigma-i\tilde t_{\sigma\lambda -}d_\sigma^\dagger \tilde f_\lambda + i\tilde t_{\sigma\lambda +} d_\sigma^\dagger \tilde f_{\lambda}^\dagger\,,
\end{align}
where $\tilde t_{\sigma\lambda \pm}=\tilde t_{\sigma\lambda}'\pm \tilde t_{\sigma\lambda}$.
Using a Schrieffer-Wolff transformation\cite{schriefferPR66,bravyiAoP11}, one may show that the operators $A_{\sigma\lambda}-A_{\sigma\lambda}^\dagger$ and $B_{\sigma\lambda}-B_{\sigma\lambda}^\dagger$ eliminate the tunneling Hamiltonian, $\tilde H_T=-\sum_{\sigma\lambda}[A_{\sigma\lambda}-A_{\sigma\lambda}^\dagger+B_{\sigma\lambda}-B_{\sigma\lambda}^\dagger,\tilde H_M+H_D]$, to first order in $\tilde t_{\sigma\lambda\pm}$, where
\begin{align}
A_{\sigma\lambda}&=i(\tilde t_{\sigma\lambda}^*-\tilde t_{\sigma\lambda}'^*) \left[\frac{1}{\epsilon_\sigma-2\tilde\delta_\lambda} - \frac{U n_{\bar\sigma}}{(\epsilon_\sigma-2\tilde\delta_\lambda)(\epsilon_\sigma+U-2\tilde\delta_\lambda)} \right]\tilde f^\dagger_\lambda d_\sigma\nonumber\\
&=-i\tilde t_{\sigma\lambda -}^* \left[\frac{1}{\epsilon_\sigma-2\tilde\delta_\lambda} - \frac{U n_{\bar\sigma}}{(\epsilon_\sigma-2\tilde\delta_\lambda)(\epsilon_\sigma+U-2\tilde\delta_\lambda)} \right]\tilde f^\dagger_\lambda d_\sigma\,,\nonumber\\
B_{\sigma\lambda}&=i(\tilde t_{\sigma\lambda}^*+\tilde t_{\sigma\lambda}'^*)\left[\frac{1}{\epsilon_\sigma+2\tilde\delta_\lambda} - \frac{U n_{\bar\sigma}}{(\epsilon_\sigma+2\tilde\delta_\lambda)(\epsilon_\sigma+U+2\tilde\delta_\lambda)} \right]\tilde f_\lambda d_\sigma\nonumber\\
&=i\tilde t_{\sigma\lambda +}^* \left[\frac{1}{\epsilon_\sigma+2\tilde\delta_\lambda} - \frac{U n_{\bar\sigma}}{(\epsilon_\sigma+2\tilde\delta_\lambda)(\epsilon_\sigma+U+2\tilde\delta_\lambda)} \right]\tilde f_\lambda d_\sigma\,.\nonumber\\
\end{align}

We must now calculate $[A_{\rho\lambda},\tilde H_T]$ and $[B_{\rho\lambda},\tilde H_T]$, involving the commutation relations
\begin{align}
[\tilde f_\lambda^\dagger d_\rho,\tilde H_T]&=i\sum_{\sigma\kappa}[\tilde f_{\lambda}^\dagger d_\rho,\tilde t_{\sigma\kappa-}^*\tilde f^\dagger_\kappa d_\sigma - \tilde t_{\sigma\kappa+}^*\tilde f_{\kappa} d_\sigma-\tilde t_{\sigma\kappa-}d_\sigma^\dagger \tilde f_\kappa + \tilde t_{\sigma\kappa+} d_\sigma^\dagger \tilde f_{\kappa}^\dagger]\nonumber\\
&=i\sum_{\sigma\kappa}\tilde\delta_{\kappa\lambda}\tilde t_{\sigma\kappa+}^*d_\rho d_\sigma-\tilde t_{\sigma\kappa-}(\tilde\delta_{\rho\sigma}\tilde f_\lambda^\dagger \tilde f_\kappa-\tilde\delta_{\lambda\kappa}d^\dagger_\sigma d_\rho)+\tilde t_{\sigma\kappa+}\tilde\delta_{\rho\sigma}\tilde f^\dagger_\lambda \tilde f^\dagger_\kappa\,,\nonumber\\
[\tilde f_\lambda d_\rho, \tilde H_T]&=i\sum_{\sigma\kappa}[\tilde f_\lambda d_\rho,\tilde t_{\sigma\kappa-}^*\tilde f^\dagger_\kappa d_\sigma - \tilde t_{\sigma\kappa+}^*\tilde f_{\kappa} d_\sigma-\tilde t_{\sigma\kappa-}d_\sigma^\dagger \tilde f_\kappa + \tilde t_{\sigma\kappa+} d_\sigma^\dagger \tilde f_{\kappa}^\dagger]\nonumber\\
&=i\sum_{\sigma\kappa}-\tilde t_{\sigma\kappa-}^*\tilde\delta_{\kappa\lambda}d_\rho d_\sigma-\tilde t_{\sigma\kappa-}\tilde\delta_{\rho\sigma}\tilde f_\lambda \tilde f_\kappa+\tilde t_{\sigma\kappa+}(\tilde\delta_{\rho\sigma}\tilde f_\lambda \tilde f_\kappa^\dagger-\tilde\delta_{\kappa\lambda}d^\dagger_\sigma d_\rho)\,.
\end{align}
Note that $[ Un_{\bar\rho}\tilde f_\lambda^\dagger d_\rho,\tilde H_T]= Un_{\bar\rho}[\tilde f_\lambda^\dagger d_\rho,\tilde H_T]+[Un_{\bar\rho},\tilde H_T]\tilde f_\lambda^\dagger d_\rho$ and
\begin{align}
[n_{\bar\rho},\tilde H_T]&=i\sum_{\sigma\lambda}[n_{\bar\rho},\tilde t_{\sigma\lambda -}^*\tilde f^\dagger_\lambda d_\sigma - \tilde t_{\sigma\lambda +}^*\tilde f_{\lambda} d_\sigma-\tilde t_{\sigma\lambda -}d_\sigma^\dagger \tilde f_\lambda + \tilde t_{\sigma\lambda +} d_\sigma^\dagger \tilde f_{\lambda}^\dagger]\nonumber\\
&=i\sum_{\sigma\lambda}\tilde t^*_{\sigma\lambda -}\tilde\delta_{\bar\rho\sigma}d_{\bar\rho}\tilde f_\lambda^\dagger-\tilde t_{\sigma\lambda +}^*\tilde\delta_{\bar\rho\sigma}d_{\bar\rho}\tilde f_\lambda-\tilde t_{\sigma\lambda -}\tilde\delta_{\bar\rho\sigma}d_\sigma^\dagger \tilde f_\lambda+\tilde t_{\sigma\lambda +}\tilde\delta_{\bar\rho\sigma}d^\dagger_\sigma \tilde f_\lambda^\dagger\,.
\end{align}
Taking the large on-site charging limit, $U\rightarrow \infty$, we find
\begin{align}
\sum_{\rho\lambda}[A_{\rho\lambda},\tilde H_T]&=-i\sum_{\rho\lambda} \tilde t_{\rho\lambda -}^*\left[\left(\frac{1}{\epsilon_\rho-2\tilde\delta_\lambda} - \frac{ n_{\bar\rho}}{\epsilon_\rho-2\tilde\delta_\lambda} \right)[\tilde f^\dagger_\lambda d_\rho,\tilde H_T]-\frac{ [n_{\bar\rho},\tilde H_T]\tilde f^\dagger_\lambda d_\rho}{\epsilon_\rho-2\tilde\delta_\lambda}\right]\nonumber\\
&=-i\sum_{\rho\lambda}\frac{\tilde t_{\rho\lambda -}^*}{\epsilon_\rho-2\tilde\delta_\lambda}\left[ n_\rho[\tilde f^\dagger_\lambda d_\rho,\tilde H_T]- [n_{\bar\rho},\tilde H_T]\tilde f^\dagger_\lambda d_\rho\right]\nonumber\\
&=\sum_{\sigma\rho\kappa\lambda}\frac{\tilde t_{\rho\lambda -}^*}{\epsilon_\rho-2\tilde\delta_\lambda}\left[ n_\rho(\tilde t_{\sigma\kappa+}^*\tilde\delta_{\kappa\lambda}d_\rho d_\sigma-\tilde t_{\sigma\kappa-}(\tilde\delta_{\rho\sigma}\tilde f_\lambda^\dagger \tilde f_\kappa-\tilde\delta_{\kappa\lambda}d^\dagger_\sigma d_\rho)+\tilde t_{\sigma\kappa+}\tilde\delta_{\rho\sigma}\tilde f^\dagger_\lambda \tilde f^\dagger_\kappa)\right.\nonumber\\
&\left.- (\tilde t^*_{\sigma\kappa-}\tilde\delta_{\bar\rho\sigma}d_{\bar\rho}\tilde f_\kappa^\dagger-\tilde t_{\sigma\kappa+}^*\tilde\delta_{\bar\rho\sigma}d_{\bar\rho}\tilde f_\kappa-\tilde t_{\sigma\kappa-}\tilde\delta_{\bar\rho\sigma}d_\sigma^\dagger \tilde f_\kappa+\tilde t_{\sigma\kappa+}\tilde\delta_{\bar\rho\sigma}d^\dagger_\sigma \tilde f_\kappa^\dagger)\tilde f^\dagger_\lambda d_\rho\right]\,,\nonumber\\
\sum_{\rho\lambda}[B_{\rho\lambda},\tilde H_T]&=i\sum_{\rho\lambda}\tilde t_{\rho\lambda +}^*\left[ \left(\frac{1}{\epsilon_\rho+2\tilde\delta_\lambda} - \frac{ n_{\bar\rho}}{\epsilon_\rho+2\tilde\delta_\lambda} \right)[\tilde f_\lambda d_\rho,\tilde H_T]-\frac{ [n_{\bar\rho},\tilde H_T]\tilde f_\lambda d_\rho}{\epsilon_\rho+2\tilde\delta_\lambda}\right]\nonumber\\
&=i\sum_{\rho\lambda}\frac{\tilde t_{\rho\lambda +}^*}{\epsilon_\rho+2\tilde\delta_\lambda}\left[ n_\rho[\tilde f_\lambda d_\rho,\tilde H_T]- [n_{\bar\rho},\tilde H_T]\tilde f_\lambda d_\rho\right]\nonumber\\
&=-\sum_{\sigma\rho\kappa\lambda}\frac{\tilde t_{\rho\lambda +}^*}{\epsilon_\rho+2\tilde\delta_\lambda}\left[ n_\rho(-\tilde t_{\sigma\kappa-}^*\tilde\delta_{\kappa\lambda}d_\rho d_\sigma-\tilde t_{\sigma\kappa-}\tilde\delta_{\rho\sigma}\tilde f_\lambda \tilde f_\kappa+\tilde t_{\sigma\kappa+}(\tilde\delta_{\rho\sigma}\tilde f_\lambda \tilde f_\kappa^\dagger-\tilde\delta_{\kappa\lambda}d^\dagger_\sigma d_\rho))\right.\nonumber\\
&\left.- (\tilde t^*_{\sigma\kappa-}\tilde\delta_{\bar\rho\sigma}d_{\bar\rho}\tilde f_\kappa^\dagger-\tilde t_{\sigma\kappa+}^*\tilde\delta_{\bar\rho\sigma}d_{\bar\rho}\tilde f_\kappa-\tilde t_{\sigma\kappa-}\tilde\delta_{\bar\rho\sigma}d_\sigma^\dagger \tilde f_\kappa+\tilde t_{\sigma\kappa+}\tilde\delta_{\bar\rho\sigma}d^\dagger_\sigma \tilde f_\kappa^\dagger)\tilde f_\lambda d_\rho\right]\,.\nonumber\\
\end{align}
Notice that, for $\hat O=\tilde f^\dagger_\lambda,\, \tilde f_\lambda$, $n_{\rho}[\hat O d_\rho,\tilde H_T]=-n_{\rho}\tilde H_T\hat O d_\rho$. The only term that survives from $\tilde H_T$ is proportional to $d_{\rho}^\dagger$ so that this term has no spin flip processes:
\begin{align}
-n_\rho \tilde H_T \tilde f_\lambda^\dagger d_\rho&=i( \tilde t_{\rho\kappa-}n_\rho d^\dagger_\rho \tilde f_\kappa-\tilde t_{\rho\kappa+}d^\dagger_\rho \tilde f_\kappa^\dagger) \tilde f_\lambda^\dagger d_\rho=i( \tilde t_{\rho\kappa-} \tilde f_\kappa \tilde f_\lambda^\dagger-\tilde t_{\rho\kappa+} \tilde f_\kappa^\dagger \tilde f_\lambda^\dagger)n_\rho\,,\nonumber\\
-n_\rho \tilde H_T \tilde f_\lambda d_\rho&=-i(-\tilde t_{\rho\kappa+}n_\rho d^\dagger_\rho \tilde f_\kappa^\dagger+\tilde t_{\rho\kappa-}d_\rho^\dagger \tilde f_\kappa)  \tilde f_\lambda d_\rho=-i(-\tilde t_{\rho\kappa+}\tilde f_\kappa^\dagger \tilde f_\lambda +\tilde t_{\rho\kappa-} \tilde f_\kappa \tilde f_\lambda)n_\rho\,.
\end{align}
Therefore, these terms do not involve spin flips and
\begin{align}
\sum_{\rho\lambda}[A_{\rho\lambda},\tilde H_T]&=-i\sum_{\rho\lambda}\tilde t_{\rho\lambda -}^*\left[ \left(\frac{1}{\epsilon_\rho-2\tilde\delta_\lambda} - \frac{ n_{\bar\rho}}{\epsilon_\rho-2\tilde\delta_\lambda} \right)[\tilde f^\dagger_\lambda d_\rho,\tilde H_T]-\frac{ [n_{\bar\rho},\tilde H_T]\tilde f^\dagger_\lambda d_\rho}{\epsilon_\rho-2\tilde\delta_\lambda}\right]\nonumber\\
&=-i\sum_{\rho\lambda}\frac{\tilde t_{\rho\lambda -}^*}{\epsilon_\rho-2\tilde\delta_\lambda}\left[ n_\rho[\tilde f^\dagger_\lambda d_\rho,\tilde H_T]- [n_{\bar\rho},\tilde H_T]\tilde f^\dagger_\lambda d_\rho\right]\nonumber\\
&=\sum_{\sigma\rho\kappa\lambda}\frac{\tilde t_{\rho\lambda -}^*}{\epsilon_\rho-2\tilde\delta_\lambda}\left[ (\tilde t_{\sigma\kappa-}\tilde f_\kappa \tilde f_\lambda^\dagger-\tilde t_{\sigma\kappa+}\tilde f_\kappa^\dagger \tilde f_\lambda^\dagger)\tilde\delta_{\sigma\rho}n_\rho \right. \nonumber\\
&\left.~~~~~~~~~~~~~~~~~~~~~- (\tilde t^*_{\sigma\kappa-}\tilde\delta_{\bar\rho\sigma}d_{\bar\rho}\tilde f_\kappa^\dagger-\tilde t_{\sigma\kappa+}^*\tilde\delta_{\bar\rho\sigma}d_{\bar\rho}\tilde f_\kappa-\tilde t_{\sigma\kappa-}\tilde\delta_{\bar\rho\sigma}d_\sigma^\dagger \tilde f_\kappa+\tilde t_{\sigma\kappa+}\tilde\delta_{\bar\rho\sigma}d^\dagger_\sigma \tilde f_\kappa^\dagger)\tilde f^\dagger_\lambda d_\rho\right]\nonumber\\
&=\sum_{\sigma\rho\kappa\lambda}\frac{\tilde t_{\rho\lambda -}^*}{\epsilon_\rho-2\tilde\delta_\lambda}\left[ \tilde t_{\sigma\kappa-}\tilde\delta_{\sigma\rho}n_\rho \tilde f_\kappa \tilde f_\lambda^\dagger-\tilde t_{\sigma\kappa+}\tilde\delta_{\sigma\rho}n_\rho \tilde f_\kappa^\dagger \tilde f_\lambda^\dagger- (-\tilde t_{\sigma\kappa-}\tilde\delta_{\bar\rho\sigma}d_\sigma^\dagger \tilde f_\kappa+\tilde t_{\sigma\kappa+}\tilde\delta_{\bar\rho\sigma}d^\dagger_\sigma \tilde f_\kappa^\dagger)\tilde f^\dagger_\lambda d_\rho\right]\,,\nonumber\\
\sum_{\rho\lambda}[B_{\rho\lambda},\tilde H_T]&=i\sum_{\rho\lambda} \tilde t_{\rho\lambda +}^*\left[\left(\frac{1}{\epsilon_\rho+2\tilde\delta_\lambda} - \frac{ n_{\bar\rho}}{\epsilon_\rho+2\tilde\delta_\lambda} \right)[\tilde f_\lambda d_\rho,\tilde H_T]-\frac{ [n_{\bar\rho},\tilde H_T]\tilde f_\lambda d_\rho}{\epsilon_\rho+2\tilde\delta_\lambda}\right]\nonumber\\
&=i\sum_{\rho\lambda}\frac{\tilde t_{\rho\lambda +}^*}{\epsilon_\rho+2\tilde\delta_\lambda}\left[ n_\rho[\tilde f_\lambda d_\rho,\tilde H_T]- [n_{\bar\rho},\tilde H_T]\tilde f_\lambda d_\rho\right]\nonumber\\
&=-\sum_{\sigma\rho\kappa\lambda}\frac{\tilde t_{\rho\lambda +}^*}{\epsilon_\rho+2\tilde\delta_\lambda}[-\tilde t_{\sigma\kappa+}\tilde\delta_{\sigma\rho}n_\rho \tilde f_\kappa^\dagger \tilde f_\lambda+\tilde t_{\sigma\kappa-}\tilde\delta_{\sigma\rho}n_\rho \tilde f_\kappa \tilde f_\lambda\nonumber\\
&~~~~~~~~~~~~~~~~~~~~~~~- (\tilde t^*_{\sigma\kappa-}\tilde\delta_{\bar\rho\sigma}d_{\bar\rho}\tilde f_\kappa^\dagger-\tilde t_{\sigma\kappa+}^*\tilde\delta_{\bar\rho\sigma}d_{\bar\rho}\tilde f_\kappa-\tilde t_{\sigma\kappa-}\tilde\delta_{\bar\rho\sigma}d_\sigma^\dagger \tilde f_\kappa+\tilde t_{\sigma\kappa+}\tilde\delta_{\bar\rho\sigma}d^\dagger_\sigma \tilde f_\kappa^\dagger) \tilde f_\lambda d_\rho]\nonumber\\
&=-\sum_{\sigma\rho\kappa\lambda}\frac{\tilde t_{\rho\lambda +}^*}{\epsilon_\rho+2\tilde\delta_\lambda}\left[-\tilde t_{\sigma\kappa+}\tilde\delta_{\sigma\rho}n_\rho \tilde f_\kappa^\dagger \tilde f_\lambda+\tilde t_{\sigma\kappa-}\tilde\delta_{\sigma\rho}n_\rho \tilde f_\kappa \tilde f_\lambda- (-\tilde t_{\sigma\kappa-}\tilde\delta_{\bar\rho\sigma}d_\sigma^\dagger \tilde f_\kappa+\tilde t_{\sigma\kappa+}\tilde\delta_{\bar\rho\sigma}d^\dagger_\sigma \tilde f_\kappa^\dagger)\tilde f_\lambda d_\rho\right]\,.
\end{align}
Let us consider processes when only one wire is involved in then tunneling, $\kappa=\lambda$:
\begin{align}
\sum_{\rho\lambda}[A_{\rho\lambda},\tilde H_T]&=\sum_{\sigma\rho\kappa\lambda}\frac{\tilde t_{\rho\lambda -}^*}{\epsilon_\rho-2\tilde\delta_\lambda}\left[ \tilde t_{\sigma\kappa-}\tilde\delta_{\sigma\rho}n_\rho \tilde f_\kappa \tilde f_\lambda^\dagger-\tilde t_{\sigma\kappa+}\tilde\delta_{\sigma\rho}n_\rho \tilde f_\kappa^\dagger \tilde f_\lambda^\dagger- (-\tilde t_{\sigma\kappa-}\tilde\delta_{\bar\rho\sigma}d_\sigma^\dagger \tilde f_\kappa+\tilde t_{\sigma\kappa+}\tilde\delta_{\bar\rho\sigma}d^\dagger_\sigma \tilde f_\kappa^\dagger)\tilde f^\dagger_\lambda d_\rho\right]\nonumber\\
&=\sum_{\rho\lambda}\frac{\tilde t_{\rho\lambda-}^*}{\epsilon_\rho-2\tilde\delta_\lambda}\left[ \tilde t_{\rho\lambda-}n_\rho \tilde f_\lambda \tilde f_\lambda^\dagger+\tilde t_{\bar\rho\lambda-}d_{\bar\rho}^\dagger \tilde f_\lambda \tilde f_\lambda^\dagger d_\rho\right]\nonumber\\
\sum_{\rho\lambda}[B_{\rho\lambda},\tilde H_T]&=-\sum_{\sigma\rho\kappa\lambda}\frac{\tilde t_{\rho\lambda +}^*}{\epsilon_\rho+2\tilde\delta_\lambda}\left[-\tilde t_{\sigma\kappa+}\tilde\delta_{\sigma\rho}n_\rho \tilde f_\kappa^\dagger \tilde f_\lambda+\tilde t_{\sigma\kappa-}\tilde\delta_{\sigma\rho}n_\rho \tilde f_\kappa \tilde f_\lambda- (-\tilde t_{\sigma\kappa-}\tilde\delta_{\bar\rho\sigma}d_\sigma^\dagger \tilde f_\kappa+\tilde t_{\sigma\kappa+}\tilde\delta_{\bar\rho\sigma}d^\dagger_\sigma \tilde f_\kappa^\dagger)\tilde f_\lambda d_\rho\right]\nonumber\\
&=-\sum_{\rho\lambda}\frac{\tilde t_{\rho\lambda+}^*}{\epsilon_\rho+2\tilde\delta}\left[-\tilde t_{\rho\lambda+}n_\rho \tilde f_\lambda^\dagger \tilde f_\lambda - \tilde t_{\bar\rho\lambda+}\tilde f_\lambda d_{\bar\rho}^\dagger \tilde f_\lambda^\dagger \tilde f_\lambda d_\rho\right]\nonumber\\
&=\sum_{\rho\lambda}\frac{\tilde t_{\rho\lambda+}^*}{\epsilon_\rho+2\tilde\delta}\left[\tilde t_{\rho\lambda+}n_\rho \tilde f_\lambda^\dagger \tilde f_\lambda +\tilde t_{\bar\rho\lambda+}d^\dagger_{\bar\rho} \tilde f_\lambda^\dagger \tilde f_\lambda d_\rho\right]\,,
\end{align}
Summing these together with their Hermitian conjugate, we get
\begin{align}
\tilde{\mathcal H}_s&=\sum_{\rho\lambda}2n_\rho\left(\frac{|\tilde t_{\rho\lambda+}|^2}{\epsilon_\rho+2\tilde\delta_\lambda}\tilde f_\lambda^\dagger \tilde f_\lambda +\frac{|\tilde t_{\rho\lambda-}|^2}{\epsilon_\rho-2\tilde\delta_\lambda}\tilde f_\lambda \tilde f_\lambda^\dagger\right)+d^\dagger_{\bar\rho}d_\rho\left(\frac{\tilde t^*_{\rho\lambda+}\tilde t_{\bar\rho\lambda+}}{\epsilon_\rho+2\tilde\delta_\lambda}\tilde f_\lambda^\dagger \tilde f_\lambda+\frac{\tilde t^*_{\rho\lambda-}\tilde t_{\bar\rho\lambda-}}{\epsilon_\rho-2\tilde\delta_\lambda}\tilde f_\lambda \tilde f_\lambda^\dagger\right)\nonumber\\
&+d^\dagger_{\bar\rho}d_\rho\left(\frac{\tilde t^*_{\rho\lambda+}\tilde t_{\bar\rho\lambda+}}{\epsilon_{\bar\rho}+2\tilde\delta_\lambda}\tilde f_\lambda^\dagger \tilde f_\lambda+\frac{\tilde t^*_{\rho\lambda-}\tilde t_{\bar\rho\lambda-}}{\epsilon_{\bar\rho}-2\tilde\delta_\lambda}\tilde f_\lambda \tilde f_\lambda^\dagger\right)\nonumber\\
&=\sum_{\rho\lambda}2n_\rho\left(\frac{|\tilde t_{\rho\lambda+}|^2}{\epsilon_\rho+2\tilde\delta_\lambda}\tilde f_\lambda^\dagger \tilde f_\lambda +\frac{|\tilde t_{\rho\lambda-}|^2}{\epsilon_\rho-2\tilde\delta_\lambda}\tilde f_\lambda \tilde f_\lambda^\dagger\right)+d^\dagger_{\rho}d_{\bar\rho}\left(\frac{\tilde t^*_{\bar\rho\lambda+}\tilde t_{\rho\lambda+}}{\epsilon_{\bar\rho}+2\tilde\delta_\lambda}\tilde f_\lambda^\dagger \tilde f_\lambda+\frac{\tilde t^*_{\bar\rho\lambda-}\tilde t_{\rho\lambda-}}{\epsilon_{\bar\rho}-2\tilde\delta_\lambda}\tilde f_\lambda \tilde f_\lambda^\dagger\right)\nonumber\\
&+d^\dagger_{\rho}d_{\bar\rho}\left(\frac{\tilde t^*_{\bar\rho\lambda+}\tilde t_{\rho\lambda+}}{\epsilon_{\rho}+2\tilde\delta_\lambda}\tilde f_\lambda^\dagger \tilde f_\lambda+\frac{\tilde t^*_{\bar\rho\lambda-}\tilde t_{\rho\lambda-}}{\epsilon_{\rho}-2\tilde\delta_\lambda}\tilde f_\lambda \tilde f_\lambda^\dagger\right)\nonumber\\
&=\sum_{\rho\lambda}\left[2n_\rho\left(\frac{|\tilde t_{\rho\lambda+}|^2}{\epsilon_\rho+2\tilde\delta_\lambda}-\frac{|\tilde t_{\rho\lambda-}|^2}{\epsilon_\rho-2\tilde\delta_\lambda}\right)+d^\dagger_{\rho}d_{\bar\rho}\left(\frac{\tilde t^*_{\bar\rho\lambda+}\tilde t_{\rho\lambda+}}{\epsilon_{\bar\rho}+2\tilde\delta_\lambda}+\frac{\tilde t^*_{\bar\rho\lambda+}\tilde t_{\rho\lambda+}}{\epsilon_{\rho}+2\tilde\delta_\lambda}-\frac{\tilde t^*_{\bar\rho\lambda-}\tilde t_{\rho\lambda-}}{\epsilon_{\bar\rho}-2\tilde\delta_\lambda}-\frac{\tilde t^*_{\bar\rho\lambda-}\tilde t_{\rho\lambda-}}{\epsilon_{\rho}-2\tilde\delta_\lambda}\right)\right]\tilde f^\dagger_\lambda \tilde f_\lambda\nonumber\\
&+2n_\rho \frac{|\tilde t_{\rho\lambda-}|^2}{\epsilon_\rho-2\tilde\delta_\lambda}+d^\dagger_{\rho}d_{\bar\rho}\left(\frac{\tilde t^*_{\bar\rho\lambda-}\tilde t_{\rho\lambda-}}{\epsilon_{\bar\rho}-2\tilde\delta_\lambda}+\frac{\tilde t^*_{\bar\rho\lambda-}\tilde t_{\rho\lambda-}}{\epsilon_{\rho}-2\tilde\delta_\lambda}\right)\,.
\end{align}

Processes involving two wires, $\kappa=\bar\lambda$, are calculated from
\begin{align}
\sum_{\rho\lambda}[A_{\rho\lambda},\tilde H_T]&=\sum_{\sigma\rho\kappa\lambda}\frac{\tilde t_{\rho\lambda -}^*}{\epsilon_\rho-2\tilde\delta_\lambda}\left[ \tilde t_{\sigma\kappa-}\tilde\delta_{\sigma\rho}n_\rho \tilde f_\kappa \tilde f_\lambda^\dagger-\tilde t_{\sigma\kappa+}\tilde\delta_{\sigma\rho}n_\rho \tilde f_\kappa^\dagger \tilde f_\lambda^\dagger- (-\tilde t_{\sigma\kappa-}\tilde\delta_{\bar\rho\sigma}d_\sigma^\dagger \tilde f_\kappa+\tilde t_{\sigma\kappa+}\tilde\delta_{\bar\rho\sigma}d^\dagger_\sigma \tilde f_\kappa^\dagger)\tilde f^\dagger_\lambda d_\rho\right]\nonumber\\
&=\sum_{\rho\lambda}\frac{\tilde t_{\rho\lambda -}^*}{\epsilon_\rho-2\tilde\delta_\lambda}\left[ \tilde t_{\rho\bar\lambda-}n_\rho \tilde f_{\bar\lambda} \tilde f_\lambda^\dagger-\tilde t_{\rho\bar\lambda+}n_\rho \tilde f_{\bar\lambda}^\dagger \tilde f_\lambda^\dagger- (-\tilde t_{\bar\rho\bar\lambda-}d_{\bar\rho}^\dagger \tilde f_{\bar\lambda}+\tilde t_{\bar\rho\bar\lambda+}d^\dagger_{\bar\rho} \tilde f_{\bar\lambda}^\dagger)\tilde f^\dagger_\lambda d_\rho\right]\nonumber\\
\sum_{\rho\lambda}[B_{\rho\lambda},\tilde H_T]&=-\sum_{\sigma\rho\kappa\lambda}\frac{\tilde t_{\rho\lambda +}^*}{\epsilon_\rho+2\tilde\delta_\lambda}\left[-\tilde t_{\sigma\kappa+}\tilde\delta_{\sigma\rho}n_\rho \tilde f_\kappa^\dagger \tilde f_\lambda+\tilde t_{\sigma\kappa-}\tilde\delta_{\sigma\rho}n_\rho \tilde f_\kappa \tilde f_\lambda- (-\tilde t_{\sigma\kappa-}\tilde\delta_{\bar\rho\sigma}d_\sigma^\dagger \tilde f_\kappa+\tilde t_{\sigma\kappa+}\tilde\delta_{\bar\rho\sigma}d^\dagger_\sigma \tilde f_\kappa^\dagger)\tilde f_\lambda d_\rho\right]\nonumber\\
&=-\sum_{\rho\lambda}\frac{\tilde t_{\rho\lambda +}^*}{\epsilon_\rho+2\tilde\delta_\lambda}\left[-\tilde t_{\rho\bar\lambda+}n_\rho \tilde f_{\bar\lambda}^\dagger \tilde f_\lambda+\tilde t_{\rho\bar\lambda-}n_\rho \tilde f_{\bar\lambda} \tilde f_\lambda- (-\tilde t_{\bar\rho\bar\lambda-}d_{\bar\rho}^\dagger \tilde f_{\bar\lambda}+\tilde t_{\bar\rho\bar\lambda+}d^\dagger_{\bar\rho} \tilde f_{\bar\lambda}^\dagger)\tilde f_\lambda d_\rho\right]\nonumber\\
&=\sum_{\rho\lambda}\frac{\tilde t_{\rho\lambda +}^*}{\epsilon_\rho+2\tilde\delta_\lambda}\left[\tilde t_{\rho\bar\lambda+}n_\rho \tilde f_{\bar\lambda}^\dagger \tilde f_\lambda-\tilde t_{\rho\bar\lambda-}n_\rho \tilde f_{\bar\lambda} \tilde f_\lambda+ (-\tilde t_{\bar\rho\bar\lambda-}d_{\bar\rho}^\dagger \tilde f_{\bar\lambda}+\tilde t_{\bar\rho\bar\lambda+}d^\dagger_{\bar\rho} \tilde f_{\bar\lambda}^\dagger)\tilde f_\lambda d_\rho\right],
\end{align}
Because we will have to add the Hermitian conjugates of these terms, we note that 
\begin{align}
\left[\sum_{\rho\lambda}\frac{\tilde t_{\rho\lambda -}^*}{\epsilon_\rho-2\tilde\delta_{\lambda}} \tilde t_{\rho\bar\lambda -}n_\rho \tilde f_{\bar\lambda} \tilde f_{\lambda}^\dagger\right]^\dagger&=\sum_{\rho\lambda}\frac{\tilde t_{\rho\lambda -}^*}{\epsilon_\rho-2\tilde\delta_{\bar\lambda}} \tilde t_{\rho\bar\lambda -}n_\rho \tilde f_{\bar\lambda} \tilde f_{\lambda}^\dagger\,,\nonumber\\
\left[\sum_{\rho\lambda}\frac{\tilde t_{\rho\lambda +}^*}{\epsilon_\rho+2\tilde\delta_{\lambda}} \tilde t_{\rho\bar\lambda +}n_\rho \tilde f_{\bar\lambda}^\dagger \tilde f_{\lambda}\right]^\dagger&=\sum_{\rho\lambda}\frac{\tilde t_{\rho\lambda +}^*}{\epsilon_\rho+2\tilde\delta_{\bar\lambda}} \tilde t_{\rho\bar\lambda +}n_\rho \tilde f_{\bar\lambda}^\dagger \tilde f_{\lambda}\,,\nonumber\\
\left[\sum_{\rho\lambda}\frac{\tilde t_{\rho\lambda -}^*}{\epsilon_\rho-2\tilde\delta_{\lambda}} \tilde t_{\bar\rho\bar\lambda -}d_{\bar\rho}^\dagger d_\rho \tilde f_{\bar\lambda} \tilde f_{\lambda}^\dagger\right]^\dagger&=\sum_{\rho\lambda}\frac{\tilde t_{\rho\lambda -}^*}{\epsilon_{\bar\rho}-2\tilde\delta_{\bar\lambda}} \tilde t_{\bar\rho\bar\lambda -}d_{\bar\rho}^\dagger d_\rho  \tilde f_{\bar\lambda} \tilde f_{\lambda}^\dagger\,,\nonumber\\
\left[\sum_{\rho\lambda}\frac{\tilde t_{\rho\lambda +}^*}{\epsilon_\rho+2\tilde\delta_{\lambda}} \tilde t_{\bar\rho\bar\lambda +}d_{\bar\rho}^\dagger d_\rho \tilde f_{\bar\lambda}^\dagger \tilde f_{\lambda}\right]^\dagger&=\sum_{\rho\lambda}\frac{\tilde t_{\rho\lambda +}^*}{\epsilon_{\bar\rho}+2\tilde\delta_{\bar\lambda}} \tilde t_{\bar\rho\bar\lambda +}d_{\bar\rho}^\dagger d_\rho \tilde f_{\bar\lambda}^\dagger \tilde f_{\lambda}\,,\nonumber\\
\end{align}
so that the contribution from the transfer of the fermions [Fig.~2(c),~(d)] is
\begin{align}
 \tilde{\mathcal H}_o=\sum_{\rho\lambda}&\left[\left(\frac{1}{{\epsilon_\rho-2\tilde\delta_{\bar\lambda}}}+\frac{1}{{\epsilon_\rho-2\tilde\delta_{\lambda}}}\right)\tilde t_{\rho\lambda -}\tilde t_{\rho\bar\lambda -}^*\tilde f_\lambda \tilde f^\dagger_{\bar\lambda}+\left(\frac{1}{{\epsilon_\rho+2\tilde\delta_{\bar\lambda}}}+\frac{1}{{\epsilon_\rho+2\tilde\delta_{\lambda}}}\right)\tilde t_{\rho\lambda +}\tilde t_{\rho\bar\lambda +}^*\tilde f^\dagger_\lambda \tilde f_{\bar\lambda}\right]n_\rho\nonumber\\
&+\left[\left(\frac{1}{\epsilon_\rho-2\tilde\delta_{\bar\lambda}}+\frac{1}{\epsilon_{\bar\rho}-2\tilde\delta_{\lambda}}\right)\tilde t^*_{\rho\bar\lambda -}\tilde t_{\bar\rho\lambda -}\tilde f_\lambda \tilde f_{\bar\lambda}^\dagger+\left(\frac{1}{\epsilon_\rho+2\tilde\delta_{\bar\lambda}}+\frac{1}{\epsilon_{\bar\rho}+2\tilde\delta_{\lambda}}\right)\tilde t^*_{\rho\bar\lambda +}\tilde t_{\bar\rho\lambda +} \tilde f_\lambda^\dagger \tilde f_{\bar\lambda}\right]d^\dagger_{\bar\rho}d_\rho\nonumber\\
=\sum_{\rho\lambda}&\left[\left(\frac{1}{{\epsilon_\rho-2\tilde\delta_{\lambda}}}+\frac{1}{{\epsilon_\rho-2\tilde\delta_{\bar\lambda}}}\right)\tilde t_{\rho\bar\lambda -}\tilde t_{\rho\lambda -}^*\tilde f_{\bar\lambda} \tilde f^\dagger_{\lambda}+\left(\frac{1}{{\epsilon_\rho+2\tilde\delta_{\bar\lambda}}}+\frac{1}{{\epsilon_\rho+2\tilde\delta_{\lambda}}}\right)\tilde t_{\rho\lambda +}\tilde t_{\rho\bar\lambda +}^*\tilde f^\dagger_\lambda \tilde f_{\bar\lambda}\right]n_\rho\nonumber\\
&+\left[\left(\frac{1}{\epsilon_\rho-2\tilde\delta_{\lambda}}+\frac{1}{\epsilon_{\bar\rho}-2\tilde\delta_{\bar\lambda}}\right)\tilde t^*_{\rho\lambda -}\tilde t_{\bar\rho\bar\lambda -}\tilde f_{\bar\lambda} \tilde f_{\lambda}^\dagger+\left(\frac{1}{\epsilon_\rho+2\tilde\delta_{\bar\lambda}}+\frac{1}{\epsilon_{\bar\rho}+2\tilde\delta_{\lambda}}\right)\tilde t^*_{\rho\bar\lambda +}\tilde t_{\bar\rho\lambda +} \tilde f_\lambda^\dagger \tilde f_{\bar\lambda}\right]d^\dagger_{\bar\rho}d_\rho\nonumber\\
=\sum_{\rho\lambda}&\left\{\left[\left(\frac{1}{{\epsilon_\rho+2\tilde\delta_{\bar\lambda}}}+\frac{1}{{\epsilon_\rho+2\tilde\delta_{\lambda}}}\right)\tilde t_{\rho\lambda +}\tilde t_{\rho\bar\lambda +}^*-\left(\frac{1}{{\epsilon_\rho-2\tilde\delta_{\lambda}}}+\frac{1}{{\epsilon_\rho-2\tilde\delta_{\bar\lambda}}}\right)\tilde t_{\rho\bar\lambda -}\tilde t_{\rho\lambda -}^*\right]\right.n_\rho\nonumber\\
&+\left.\left[\left(\frac{1}{\epsilon_\rho+2\tilde\delta_{\bar\lambda}}+\frac{1}{\epsilon_{\bar\rho}+2\tilde\delta_{\lambda}}\right)\tilde t^*_{\rho\bar\lambda +}\tilde t_{\bar\rho\lambda +}-\left(\frac{1}{\epsilon_\rho-2\tilde\delta_{\lambda}}+\frac{1}{\epsilon_{\bar\rho}-2\tilde\delta_{\bar\lambda}}\right)\tilde t^*_{\rho\lambda -}\tilde t_{\bar\rho\bar\lambda -}\right]d^\dagger_{\bar\rho}d_\rho\right\}\tilde f^\dagger_\lambda \tilde f_{\bar\lambda}\nonumber\\
=\sum_{\rho\lambda}&\left\{\left[\left(\frac{1}{{\epsilon_\rho+2\tilde\delta_{\bar\lambda}}}+\frac{1}{{\epsilon_\rho+2\tilde\delta_{\lambda}}}\right)\tilde t_{\rho\lambda +}\tilde t_{\rho\bar\lambda +}^*-\left(\frac{1}{{\epsilon_\rho-2\tilde\delta_{\lambda}}}+\frac{1}{{\epsilon_\rho-2\tilde\delta_{\bar\lambda}}}\right)\tilde t_{\rho\bar\lambda -}\tilde t_{\rho\lambda -}^*\right]\right.n_\rho\nonumber\\
&+\left.\left[\left(\frac{1}{\epsilon_{\bar\rho}+2\tilde\delta_{\bar\lambda}}+\frac{1}{\epsilon_{\rho}+2\tilde\delta_{\lambda}}\right)\tilde t^*_{\bar\rho\bar\lambda +}\tilde t_{\rho\lambda +}-\left(\frac{1}{\epsilon_{\bar\rho}-2\tilde\delta_{\lambda}}+\frac{1}{\epsilon_{\rho}-2\tilde\delta_{\bar\lambda}}\right)\tilde t^*_{\bar\rho\lambda -}\tilde t_{\rho\bar\lambda -}\right]d^\dagger_{\rho}d_{\bar\rho}\right\}\tilde f^\dagger_\lambda \tilde f_{\bar\lambda}\,.
\end{align}

Instead of forming Dirac fermions in the same wire, one can instead form a full fermion from the MFs closest together (inner fermion) and a fermion from the MFs furthest apart (outer fermion) as in the main text, $f_r=(\gamma_{r}'+i\gamma_{l})/2$ and $ f_l=(\gamma_{l}'+i\gamma_{r})/2$, respectively. The MFs are, in turn, written as $\gamma_{\lambda}'= f_\lambda+ f_\lambda^\dagger$ and $\gamma_{\lambda}=( f_{\bar\lambda}- f_{\bar\lambda}^\dagger)/i$.

The tunneling Hamiltonian can then be written as
\begin{align}
 H_T&=\sum_{\sigma,\lambda}i t_{\sigma\lambda}'d_\sigma^\dagger(f_\lambda+f_\lambda^\dagger)-i t_{\sigma\lambda}d_\sigma^\dagger(f_{\bar\lambda}-f_{\bar\lambda}^\dagger)-i  t_{\sigma\lambda}^*(f_{\bar\lambda}-f_{\bar\lambda}^\dagger)d_\sigma-i{ t_{\sigma\lambda}'}(f_\lambda+f_\lambda^\dagger)d_\sigma\nonumber\\
&=\sum_{\sigma,\lambda}i d_\sigma^\dagger[( t_{\sigma\lambda}'- t_{\bar\lambda}) f_\lambda+( t_{\sigma\lambda}'+ t_{\sigma\bar\lambda})f_{\lambda}^\dagger]-i[({ t'^*_{\sigma\lambda}}+ t_{\sigma\bar\lambda}^*)f_\lambda+({ t'^*_{\sigma\lambda}}- t_{\sigma\bar\lambda}^*)f_\lambda^\dagger]d_\sigma\nonumber\\
&=\sum_{\sigma,\lambda}-i t_{\sigma\lambda -}d_\sigma^\dagger f_\lambda+i  t_{\sigma\lambda -}^* f_{\lambda}^\dagger d_\sigma + i  t_{\sigma\lambda +} d_\sigma^\dagger f_\lambda^\dagger-i  t_{\sigma\lambda +}^* f_\lambda d_\sigma\,,
\end{align}
where we have defined $  t_{\sigma\lambda\pm}= t_{\sigma\bar\lambda}\pm  t_{\sigma\lambda}'$. Furthermore, we redefine the MF coupling in the wire so that $ H_M=\sum_\lambda\delta_\lambda(2f_\lambda^\dagger f_\lambda-1)$ where $\delta_r$ ($\delta_l$) now parameterizes the overlap between the inner (outer) MFs. With this redefinition, we see that the transformed Hamiltonian is, term by term, identical to Eq.~({2}) with the exchange of tilded to untilded variables. Therefore, upon performing the same Schrieffer-Wolff transformation we find
\begin{align}
\mathcal H_s&=\sum_{\rho\lambda}\left[2n_\rho\left(\frac{| t_{\rho\lambda+}|^2}{\epsilon_\rho+2\delta_\lambda}-\frac{| t_{\rho\lambda-}|^2}{\epsilon_\rho-2\delta_\lambda}\right)+d^\dagger_{\rho}d_{\bar\rho}\left(\frac{ t^*_{\bar\rho\lambda+} t_{\rho\lambda+}}{\epsilon_{\bar\rho}+2\delta_\lambda}+\frac{ t^*_{\bar\rho\lambda+} t_{\rho\lambda+}}{\epsilon_{\rho}+2\delta_\lambda}-\frac{ t^*_{\bar\rho\lambda-} t_{\rho\lambda-}}{\epsilon_{\bar\rho}-2\delta_\lambda}-\frac{ t^*_{\bar\rho\lambda-} t_{\rho\lambda-}}{\epsilon_{\rho}-2\delta_\lambda}\right)\right] f^\dagger_\lambda  f_\lambda\nonumber\\
&+2n_\rho \frac{| t_{\rho\lambda-}|^2}{\epsilon_\rho-2\delta_\lambda}+d^\dagger_{\rho}d_{\bar\rho}\left(\frac{ t^*_{\bar\rho\lambda-} t_{\rho\lambda-}}{\epsilon_{\bar\rho}-2\delta_\lambda}+\frac{ t^*_{\bar\rho\lambda-} t_{\rho\lambda-}}{\epsilon_{\rho}-2\delta_\lambda}\right)\,,\\
\mathcal H_o&=\sum_{\rho\lambda}\left\{\left[\left(\frac{1}{{\epsilon_\rho+2\delta_{\bar\lambda}}}+\frac{1}{{\epsilon_\rho+2\delta_{\lambda}}}\right) t_{\rho\lambda +} t_{\rho\bar\lambda +}^*-\left(\frac{1}{{\epsilon_\rho-2\delta_{\lambda}}}+\frac{1}{{\epsilon_\rho-2\delta_{\bar\lambda}}}\right) t_{\rho\bar\lambda -} t_{\rho\lambda -}^*\right]\right.n_\rho\nonumber\\
&+\left.\left[\left(\frac{1}{\epsilon_{\bar\rho}+2\delta_{\bar\lambda}}+\frac{1}{\epsilon_{\rho}+2\delta_{\lambda}}\right) t^*_{\bar\rho\bar\lambda +} t_{\rho\lambda +}-\left(\frac{1}{\epsilon_{\bar\rho}-2\delta_{\lambda}}+\frac{1}{\epsilon_{\rho}-2\delta_{\bar\lambda}}\right) t^*_{\bar\rho\lambda -} t_{\rho\bar\lambda -}\right]d^\dagger_{\rho}d_{\bar\rho}\right\} f^\dagger_\lambda  f_{\bar\lambda}\,.
\end{align}
In the case considered in the main text, we consider coupling only to the inner MFs, so that $t_{\sigma l}'=t_{\sigma r}=0$, $t_{\sigma l\pm=0}$ and thus $\mathcal H_o=0$ and
\begin{align}
\mathcal H_s&=\sum_{\rho}\left[2n_\rho\frac{| t_{\rho+}|^2}{\epsilon_\rho+2\delta}+d^\dagger_{\rho}d_{\bar\rho}\left(\frac{ t^*_{\bar\rho+} t_{\rho+}}{\epsilon_{\bar\rho}+2\delta}+\frac{ t^*_{\bar\rho+} t_{\rho+}}{\epsilon_{\rho}+2\delta}\right)\right] f^\dagger f +\left[2n_\rho \frac{| t_{\rho-}|^2}{\epsilon_\rho-2\delta}+d^\dagger_{\rho}d_{\bar\rho}\left(\frac{ t^*_{\bar\rho-} t_{\rho-}}{\epsilon_{\bar\rho}-2\delta}+\frac{ t^*_{\bar\rho-} t_{\rho-}}{\epsilon_{\rho}-2\delta}\right)\right]ff^\dagger\,,
\end{align}
where $t_{\sigma\pm}\equiv t_{\sigma r\pm}$, $\delta\equiv\delta_r$, and $f\equiv f_r$. Performing the summation in spin, we find
\begin{align}
&\sum_\rho 2n_\rho\frac{|t_{\rho\pm}|^2}{\epsilon_\rho\pm2\delta}=2n_\uparrow\frac{|t_{\uparrow\pm}|^2}{\epsilon_\uparrow\pm2\delta}\pm2n_\downarrow\frac{|t_{\downarrow\pm}|^2}{\epsilon_\downarrow\pm2\delta}=(S_0+S_3)\frac{|t_{\uparrow\pm}|^2}{\epsilon_\uparrow\pm2\delta}+(S_0-S_3)\frac{|t_{\downarrow\pm}|^2}{\epsilon_\downarrow\pm2\delta}\nonumber\\
&=S_0\left(\frac{|t_{\uparrow\pm}|^2}{\epsilon_\uparrow\pm2\delta}+\frac{|t_{\downarrow\pm}|^2}{\epsilon_\downarrow\pm2\delta}\right)+S_3\left(\frac{|t_{\uparrow\pm}|^2}{\epsilon_\uparrow\pm2\delta}-\frac{|t_{\downarrow\pm}|^2}{\epsilon_\downarrow\pm2\delta}\right)\equiv S_0 B^\pm_0+ S_3 B^\pm_3\,,\nonumber\\
&\sum_\rho d^\dagger_{\rho}d_{\bar\rho}\left(\frac{ t^*_{\bar\rho+} t_{\rho+}}{\epsilon_{\bar\rho}\pm2\delta}+\frac{ t^*_{\bar\rho+} t_{\rho+}}{\epsilon_{\rho}\pm2\delta}\right)=d^\dagger_{\uparrow}d_{\downarrow}\left(\frac{ t^*_{\downarrow\pm} t_{\uparrow\pm}}{\epsilon_{\downarrow}\pm2\delta}+\frac{ t^*_{\downarrow\pm} t_{\uparrow\pm}}{\epsilon_{\uparrow}\pm2\delta}\right)+d^\dagger_{\downarrow}d_{\uparrow}\left(\frac{ t^*_{\uparrow\pm} t_{\downarrow\pm}}{\epsilon_{\uparrow}\pm2\delta}+\frac{ t^*_{\uparrow\pm} t_{\downarrow\pm}}{\epsilon_{\downarrow}\pm2\delta}\right)\nonumber\\
&=\frac{S_1+i S_2}{2}\left(\frac{ t^*_{\downarrow\pm} t_{\uparrow\pm}}{\epsilon_{\downarrow}\pm2\delta}+\frac{ t^*_{\downarrow\pm} t_{\uparrow\pm}}{\epsilon_{\uparrow}\pm2\delta}\right)+\frac{S_1-i S_2}{2}\left(\frac{ t^*_{\uparrow\pm} t_{\downarrow\pm}}{\epsilon_{\uparrow}\pm2\delta}+\frac{ t^*_{\uparrow\pm} t_{\downarrow\pm}}{\epsilon_{\downarrow}\pm2\delta}\right)\nonumber\\
&=\frac{S_1}{2}\left( \frac{ t^*_{\downarrow\pm} t_{\uparrow\pm}}{\epsilon_{\downarrow}\pm2\delta}+\frac{ t^*_{\downarrow\pm} t_{\uparrow\pm}}{\epsilon_{\uparrow}\pm2\delta} + \frac{ t^*_{\uparrow\pm} t_{\downarrow\pm}}{\epsilon_{\uparrow}\pm2\delta}+\frac{ t^*_{\uparrow\pm} t_{\downarrow\pm}}{\epsilon_{\downarrow}\pm2\delta}\right)+i\frac{S_2}{2}\left( \frac{ t^*_{\downarrow\pm} t_{\uparrow\pm}}{\epsilon_{\downarrow}\pm2\delta}+\frac{ t^*_{\downarrow\pm} t_{\uparrow\pm}}{\epsilon_{\uparrow}\pm2\delta} -\frac{ t^*_{\uparrow\pm} t_{\downarrow\pm}}{\epsilon_{\uparrow}\pm2\delta}-\frac{ t^*_{\uparrow\pm} t_{\downarrow\pm}}{\epsilon_{\downarrow}\pm2\delta}\right)\nonumber\\
&=S_1\textrm{Re}(t^*_{\uparrow\pm} t_{\downarrow\pm})\left(\frac{1}{\epsilon_{\uparrow}\pm2\delta}+\frac{ 1}{\epsilon_{\downarrow}\pm2\delta}\right)+S_2\textrm{Im}(t^*_{\uparrow\pm} t_{\downarrow\pm})\left(\frac{1}{\epsilon_{\uparrow}\pm2\delta}+\frac{ 1}{\epsilon_{\downarrow}\pm2\delta}\right)\equiv S_1 B^\pm_1 + S_2 B^\pm_2\,.
\end{align} 
Upon identifying $t'_{\sigma r}=t_{\sigma r}/i$ and $\mathcal H_T=\mathcal H_s$, we obtain Eq.~(\ref{tun_B}) with effective magnetic field given by Eq.~(\ref{Bs}).

\section{Numerical calculation of spin on the dot}\label{num_residual}

In this section of the Appendix we plot the $x$ component of spin on the dot $S_{j,x}$ as a function of position defined by $S_{j,x}=\tilde Y^\dagger_j \hat S_x \tilde Y_j$ (in units of $\hbar/2$), with $\tilde Y_j$ the dot wave function at site $j$ for the lowest positive energy level of the dot (see Fig.~\ref{fig:Sx_dot}). In this section, the quantum dot is far away from MFs, so the only non-zero spin projection of the dot level is $S_x$, see Fig.~\ref{fig:SxSz}.
In general, the spin oscillates at a period set by the SOI. For weak magnetic fields, these oscillations are close to be symmetric around zero so that the average spin projection on the dot is almost zero [see Fig.~\ref{fig:Sx_dot} (left panel)]. For strong magnetic fields, there is  asymmetry around zero, resulting in the average spin polarization along the magnetic field, [see Fig.~\ref{fig:Sx_dot} (right panel)]. This explains the offset in $S_x$ component of the average spin of the dot shown  in Fig.~\ref{fig:SxSz} in the main text.\cite{trifPRB08}
\begin{figure}[ht]
\includegraphics[width=8cm]{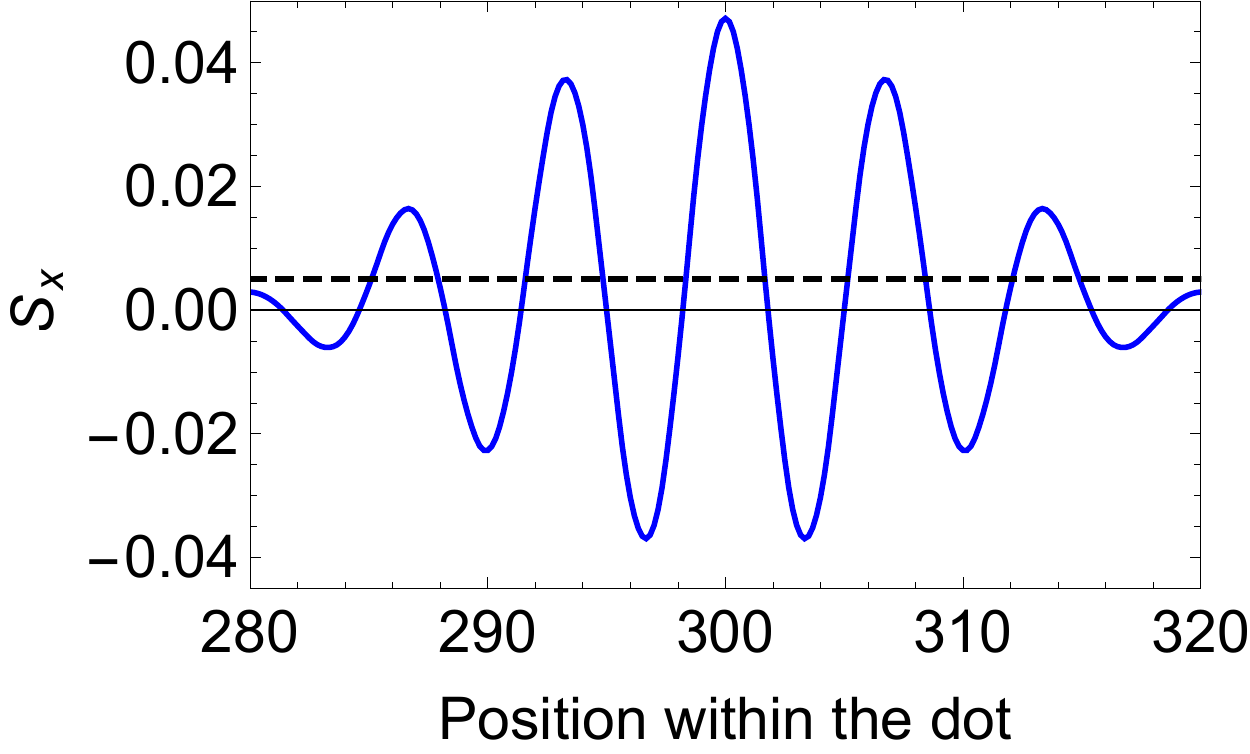}
\includegraphics[width=8cm]{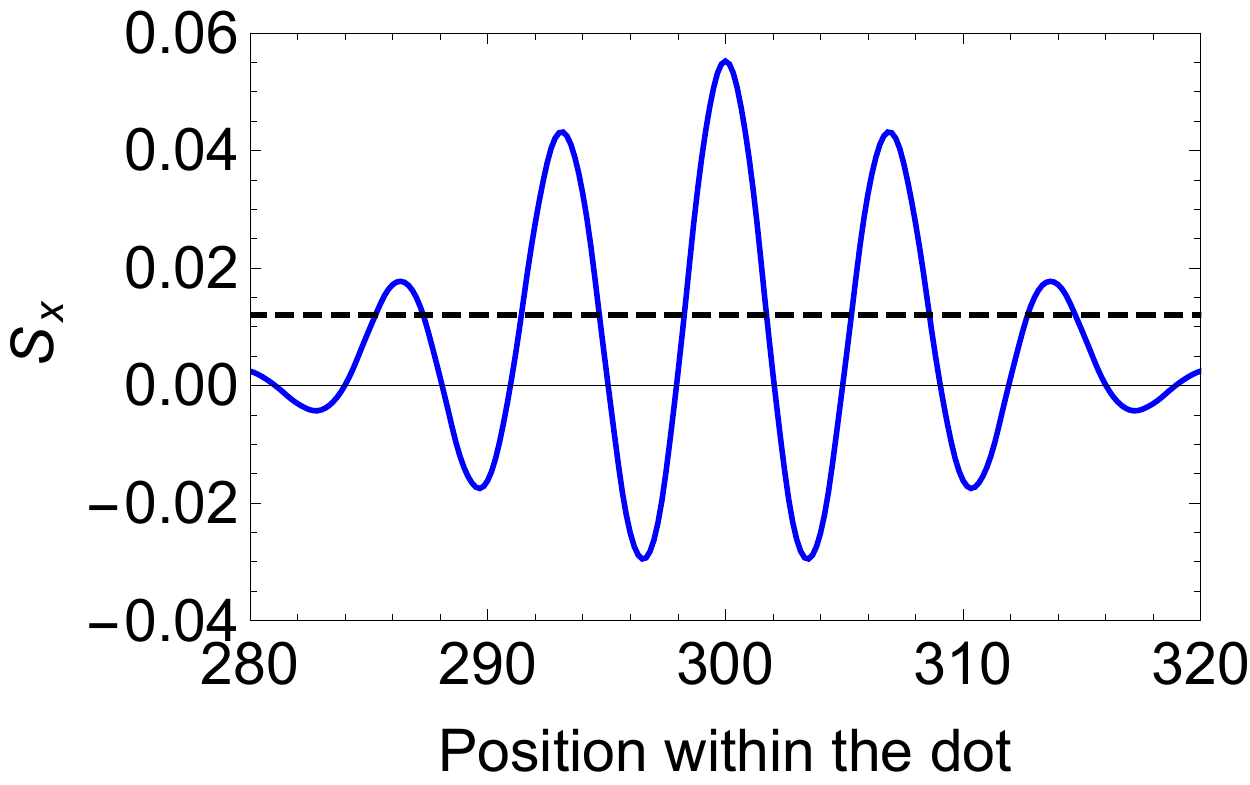}
\caption{The spin component $S_x$ of the lowest positive energy level of the dot (blue solid line) as a function of the position within the dot when the magenetic  field is 
weak ($\Delta_Z=0.04$, left panel) and strong ($\Delta_Z=0.12$, right panel). The black dashed line stands for the symmetric axis of the blue curve corresponding to the average spin projection $S_x$ on the dot. The system parameters are the same as in Fig. \ref{fig:eigenB}}.
\label{fig:Sx_dot}
\end{figure}

\end{widetext}

\end{appendix}

\end{document}